\documentclass[a4paper,fleqn,usenatbib]{mnras}

\usepackage[T1]{fontenc}
\usepackage{ae,aecompl}

\usepackage{graphicx}	
\usepackage{amsmath}	
\usepackage{amssymb}	
\usepackage{multicol}
\usepackage{bm}
\usepackage{pdflscape}

\title[Spectral calibration of K$-$M giants]{Spectral Calibration of K$-$M Giants from medium
resolution near-infrared HK-band spectra}
\author[Ghosh et al. 2018]{
Supriyo Ghosh,$^{1}$\thanks{E-mail: supriyo12a@bose.res.in (SG)}
Soumen Mondal,$^{1}$
Ramkrishna Das$^{1}$
and Dhrimadri Khata$^{1}$
\\
$^{1}$Satyendra Nath Bose National Centre for Basic Sciences,
Block-JD, Sector-III, Salt Lake, Kolkata-700 106, India\\
}

\date{Accepted 2019 January 23. Received 2019 January 23; in original form 2018 August 29}

\pubyear{2017}

\begin{document}
\label{firstpage}
\pagerange{\pageref{firstpage}--\pageref{lastpage}}
\maketitle

\begin{abstract}
We present here new medium resolution spectra ($\lambda/\Delta\lambda$ $\sim$ 1200) of K$-$M giants covering wavelength range 1.50$-$1.80 and 1.95$-$2.45 $\mu$m. The sample includes 72 K0$-$M8 giants from our TIRSPEC observations and all available 35 giants in that spectral range from archival IRTF spectral library. We have calibrated here the empirical relations between fundamental parameters (e.g., effective temperature, surface gravity) and equivalent widths of some important spectral features like Si I, Na I, Ca I, $^{12}$CO. We find that the $^{12}$CO first overtone band at 2.29 $\mu$m and second overtone band at 1.62 $\mu$m are a reasonably good indicator of temperature above 3400 K and surface gravity. We show that the dispersion of empirical relations between $^{12}$CO and $T_{eff}$ significantly improve considering the effect of surface gravity.  
\end{abstract}




\section{Introduction}
Stellar spectral libraries have a particularly important role to understand and classify the stellar population as well as an evolutionary synthesis for the individual sources in the field, star clusters of our Galaxy and integrated stellar lights in the extra-Galactic sources. For example, the original stellar classification process, developed by \citet{Morgan1943}, uses a set of reference stellar spectra to compare the spectrum of an individual star (see \citealt{Garrison1994}). Precise estimation of their fundamental parameters, e.g., effective temperature ($T_{eff}$), surface gravity (log $g$), metallicity [Fe/H], mass (M), and radius (r), from spectroscopic techniques is still challenging.

Several optical spectral libraries (e.g., ELODIE \citep{Prugniel2001},  STELIB \citep{LeBorgne2003},  CFLIB  \citep{Valdes2004}, Indo-US  \citep{Valdes2004}, MILES \citep{Sanchez2006}, CaT \citep{Cenarro2001, Cenarro2007}) are used to construct reasonable stellar population models. The near-infrared (NIR) spectral regions are more advantageous than optical as it suffers relatively less interstellar extinction, and the NIR regime allows us to probe long distance in the galaxy. NIR spectra are particularly useful for understanding the physics of cool stars like K$-$M giants (e.g., \citealt{Joyce1998, Gautschy-Loidl2004}) as these cool stars ($T_{eff} \sim 3000-6000$) emit maximum energy (peak near 1 $\mu$m) in the  NIR, which can probe the deepest regions of the stellar photosphere \citep{Forster2000}. Particularly, classifying and characterizing of individual stars in nearby embedded young clusters (e.g., \citealt{Greene1995, Peterson2008}) and optically obscured regions of the Galaxy (e.g., \citealt{Figer1995, Frogel2001, Kurtev2007, Lancon2007, Riffel2008}) are much  benefited by the use of NIR spectra.

Since the pioneering work of \citet{Johnson1970}, significant progress was made by several authors in the NIR regions (see, e.g., \citealt{Origlia1993, Wallace1997, Meyer1998} for reviews). Subsequently, several spectral libraries have been developed to construct stellar population synthesis models from NIR spectra (e.g., \citealt{Kleinmann1986, Terndrup1990, Origlia1993, Wallace1996,Wallace1997, Wallace2002, Blum1996, Joyce1998, Forster2000, Lancon2000, Ivanov2004, Marmol2008, Rayner2009, Chen2014, Feldmeier2017, Villaume2017}). Among those libraries, the NASA Infrared Telescope Facility (IRTF) spectral library offers a unique advantage of continuous coverage in the NIR and mid-IR (MIR) regime (0.8$-$5 $\mu$m) but provides limited coverage of stellar parameter range \citep{Cushing2005, Rayner2009}. The X-Shooter stellar library \citep{Chen2014} that covers optical to near-IR (0.35$-$2.5 $\mu$m) would be beneficial once it is complete. Moreover, ongoing large-scale spectroscopic surveys like, Sloan Extension for Galactic Understanding and Exploration (SEGUE; \citealt{Yanny2009}), the Radial Velocity Experiment \citep{Steinmetz2006}, the Apache Point Observatory Galactic Evolution Experiment \citep{Eisenstein2011}, the LAMOST Spectroscopic Survey of the Galactic Anti-center (LSS-GAC; \citealt{Liu2014, Yuan2015}) and Gaia \citep{Perryman2001}, will be valuable for our understanding about the formation and evolution of the Milky Way.

Despite all these efforts in the understanding of stellar population in different systems, precise estimation of fundamental parameters of cool giants still remains a challenge because of their molecular near-photospheric environment \citep{Lancon2000}. The spectral database of these cool giants is highly sparse, and more additional database would be highly valuable for their classification and characterization.  Furthermore, the understanding of quantitative diagnostic tools and quality of spectral indices have an important role to quantify the stellar absorption features. In this paper a new NIR stellar spectral library of K$-$M giants has been undertaken using the medium resolution spectra ($\lambda/\Delta\lambda$ $\sim$ 1200) covering the wavelength range 1.50$-$2.45 $\mu$m. The main motivation of the present work is to widen the existing cool stellar libraries, and more importantly, investigate how accurately the fundamental parameters (e.g., T$_{eff}$ and log $g$) can be estimated from the medium-resolution NIR HK-band spectra. In addition, the present work evaluates the systematic differences between our established relations in this paper and the existing relations in the literature derived from relatively high-resolution spectra. Particularly, the present calibration could be used to derive the fundamental parameters for relatively faint sources in the high-extinction regions from such medium-resolution spectra compared to higher-resolution spectra using big aperture telescopes. Moreover, the estimation of fundamental parameters of K- and M-giants, precisely later than M3, is still a challenging task, and none of the existing libraries contains a large sample of later M3 giants for such calibrations. The paper is organized as the details of our observations and data reduction procedures are described in section 2, section 3 presents different spectral analysis tools, and section 4 deals with our new results and discussion. Finally, the summary and conclusion of our studies are presented in section 5.

\section{Observations and Data reductions}

\begin{table*}
    \centering
	\caption{Identification of Stars observed with TIRSPEC and SpeX instruments}
	\label{Table: Identification of Stars}
	\begin{tabular}{lccccccccccr} 
    \hline
 Stars & V & ST & $T_{eff}$ & log $g$ & [Fe/H] & Parallax & Date Of & Exposure & SNR $\ddagger$  & Sky & Ref \\
 Names & mag &  & (K) & (cm/$s^2$) & (dex) &  (mas) & Observation & Time (s)$\dagger$ & & Conditions &   \\
\hline
 \bf{TIRSPEC} : &  & & & & & & & & & & \\  
 &  & & & & & & & & & & \\     
HD54810 & 4.92 & K0III & 4715 & 2.395 & -0.25 & 16.08 & 2014-12-12 & 2*(5*60) & 138 & clear sky & T1,M1 \\
HD99283 & 5.70 & K0III & 4874 & 2.476 & -0.18 & 10.63 & 2017-04-09 & 2*(3*100) & 181 & clear sky & T1,M2 \\
HD102224 & 3.72 & K0III & 4482 & 1.844 & -0.33 & 17.76 & 2015-01-17 & 2*(5*17) & 171 & clear sky & T1,M5 \\
HD69994 & 5.79 & K1III & 4571 & 2.157 & -0.07 & 5.79 & 2017-04-07 & 2*(5*80) & 144 & clear sky & T1,M2 \\
HD40657 & 4.52 & K1.5III & 4400 & 1.389 & -0.58 & 7.75 & 2014-12-12 & 2*(5*25) & 130 & clear sky & T1,M2 \\
HD85503 & 3.88 & K2III & 4504 & 2.306 & 0.25 & 26.28 & 2015-01-14 & 2*(5*40) & 116 & clear sky & T1,M1 \\
HD26846 & 4.86 & K2III & 4547 & 2.125 & 0.09 & 13.46 & 2014-12-12 & 2*(5*50) & 106 & clear sky & T1,M6 \\
HD30834 & 4.77 & K3III & 4096 & 0.925 & -0.24 & 5.42 & 2015-01-13 & 2*(5*40) & 164 & clear sky & T1,M5 \\
HD92523 & 4.99 & K3III & 4115 & 1.349 & -0.38 & 7.81 & 2015-01-17 & 2*(5*32) & 240 & clear sky & T1,M2 \\
HD97605 & 5.79 & K3III & 4606 & 2.701 & -- & 16.51 & 2017-04-09 & 2*(3*80) & 130 & clear sky & T1 \\
HD49161 & 4.77 & K4III & 4243 & 1.212 & -0.03 & 6.62 & 2014-12-12 & 2*(5*30) & 121 & clear sky & T1,M2 \\
HD70272 & 4.25 & K4+III & 3900 & 0.914 & -0.24 & 8.53 & 2015-01-17 & 2*(5*8) & 174 & clear sky & T1,M9 \\
HD99167 & 4.80 & K5III & 3865 & 1.133 & -0.06 & 8.67 & 2015-01-14 & 2*(5*30) & 153 & clear sky & T1, M10 \\
HD83787 & 5.84 & K6III & 3816 & 0.892 & -0.21 & 4.22 & 2015-01-14 & 2*(5*70) & 117 & clear sky & T1,M7 \\
HD6953 & 5.79 & K7III & 4021 & 1.662 & -- & 8.28 & 2014-12-12 & 2*(6*50) & 103 & clear sky & T1 \\
HD6966 & 6.04 & M0III & 3998 & 1.483 & -- & 6.06 & 2015-12-18 & 2*(3*40) & 144 & clear sky & T1 \\
HD18760 & 6.13 & M0III & 3605 & 0.569 & -- & 3.92 & 2016-12-20 & 2*(5*30) & 133 & clear sky & T1 \\
HD38944 & 4.74 & M0III & 3799 & 0.727 & -- & 6.24 & 2015-01-14 & 2*(5*25) & 95 & clear sky & T1 \\
HD60522 & 4.06 & M0III & 3881 & 1.110 & -0.36 & 12.04 & 2014-12-12 & 2*(5*7) & 79 & thin cloud & T1,M9 \\
HD216397 & 4.93 & M0III & 3889 & 1.352 & -- & 10.03 & 2015-08-11 & 2*(3*30) & 117 & thin cloud & T1 \\
HD7158 & 6.11 & M1III & 3747 & 0.700 & -- & 5.16 & 2015-12-18 & 2*(3*30) & 123 & clear sky & T1 \\
HD82198 & 5.37 & M1III & 3875 & 1.153 & -- & 6.80 & 2015-01-14 & 2*(3*40) & 101 & clear sky & T1 \\
HD218329 & 4.52 & M1III & 3874 & 1.123 & 0.17 & 9.92 & 2015-08-11 & 2*(3*15) & 86 & thin cloud & T1,M5 \\
HD219215 & 4.22 & M1III & 4307 & 1.749 & -- & 16.14 & 2015-08-11 & 2*(6*7) & 71 & thin cloud & T1 \\
HD119149 & 5.01 & M1.5III & 3675 & 0.714 & -- & 6.40 & 2015-01-14 & 2*(5*30) & 94 & clear sky & T1 \\
HD1013 & 4.80 & M2III & 3792 & 1.028 & -- & 8.86 & 2015-12-18 & 2*(3*7) & 131 & clear sky & T1 \\
HD33463 & 6.42 & M2III & 3491 & 0.299 & -0.05 & 3.11 & 2015-01-14 & 2*(5*30) & 115 & clear sky & T1, M3 \\
HD39732 & 7.43 & M2III & 3448 & 0.380 & -- & 2.42 & 2016-12-19 & 2*(5*25) & 111 & clear sky & T1 \\
HD43151 & 8.49 & M2III & 3335 & 0.231 & -- & 1.99 & 2016-12-19 & 2*(5*30) & 170 & clear sky & T1 \\
HD92620 & 6.02 & M2III & 3500 & -- & -- & 4.02 & 2016-12-19 & 2*(5*20) & 182 & clear sky & T1 \\
HD115521 & 4.80 & M2III & 3690 & 0.418 & -- & 4.83 & 2015-01-17 & 2*(4*7) & 142 & clear sky & T1 \\
HD16058 & 5.37 & M3III & 3572 & 0.489 & 0.08 & 5.16 & 2015-01-13 & 2*(5*20) & 94 & clear sky & T1, M11 \\
HD28168 & 8.42 & M3III & 3344 & 0.497 & -- & 2.77 & 2015-01-13 & 2*(5*60) & 83 & clear sky & T1 \\
HD66875 & 5.99 & M3III & 3509 & 0.441 & -- & 4.38 & 2016-12-19 & 2*(5*12) & 143 & clear sky & T1 \\
HD99056 & 8.79 & M3III & 3137 & 0.439 & -- & 4.52 & 2016-12-19 & 2*(5*6) & 119 & clear sky & T1 \\
HD215953 & 6.84 & M3III & 3460 & 0.823 & -- & 4.17 & 2015-08-11 & 2*(3*100) & 80 & thin cloud & T1 \\
HD223637 & 5.78 & M3III & 3622 & 0.567 & -- & 3.86 & 2015-12-19 & 2*(3*15) & 125 & clear sky & T1 \\
HD25921 & 7.10 & M3/M4III & 3522 & 0.721 & -- & 3.62 & 2016-12-20 & 2*(5*40) & 110 & clear sky & T1 \\
HD33861 & 8.64 & M3.5III & 3365 & 0.565 & -- & 2.49 & 2015-01-13 & 2*(5*70) & 115 & clear sky & T1 \\
HD224062 & 5.61 & M3/M4III & 3429 & 0.299 & -- & 5.12 & 2015-12-19 & 2*(3*5) & 93 & clear sky & T1 \\
HD5316 & 6.24 & M4III & 3481 & 0.693 & -- & 5.75 & 2015-12-18 & 2*(3*10) & 129 & clear sky & T1 \\
HD34269 & 5.65 & M4III & 3427 & 0.353 & -- & 5.79 & 2015-01-13 & 2*(5*10) & 141 & clear sky & T1 \\
HD64052 & 6.39 & M4III & 3460 & 0.783 & -- & 6.45 & 2016-12-19 & 2*(5*10) & 84 & clear sky & T1 \\
HD81028 & 6.89 & M4III & 3482 & 0.147 & -- & 2.08 & 2016-12-19 & 2*(5*20) & 135 & clear sky & T1 \\
HD206632 & 6.23 & M4III & 3367 & 0.195 & -- & 4.77 & 2015-08-11 & 2*(3*5) & 73 & thin cloud & T1 \\
HD16896 & 8.25 & M5III & 3358 & 0.360 & -- & 2.68 & 2016-12-20 & 2*(5*30) & 123 & clear sky & T1 \\
HD17491 & 6.90 & M5III & 3313 & 0.071 & -- & 3.90 & 2016-12-20 & 2*(5*6) & 107 & clear sky & T1 \\

\hline
 \end{tabular}
 V mag $-$ visual magnitude, ST $-$ spectral type, T$_{eff}$ $-$ effective temperature, log $g$ $-$ surface gravity, [Fe/H] $-$ metallicity \\
 $\dagger$Exposure Time = no. of dither position*(no. of frame in each dither position* integration time) \\
 $\ddagger$ SNR $-$ signal to noise ratio, are estimated by the method as in \citet{Stoehr2008} considering the whole H-band.  \\
 Ref $-$ references of $T_{eff}$, log $g$ and [Fe/H].
 T corresponds to the reference of $T_{eff}$ and log $g$; M corresponds to the reference of [Fe/H] \\
 Ref : (T1) \citet{McDonald2017}; (T2) \citet{McDonald2012} ; (T3) \citet{Wright2003}; (T4, M4) \citet{Cesetti2013} ; (T5, M5) \citet{Prugniel2011}; \\
 (M1) \citet{Jofre2015}; (M2) \citet{Soubiran2016}; (M3) \citet{Ho2017}; (M6) \citet{Massarotti2008}; (M7) \citet{Wu2011}; (M8) \citet{McWilliam1990}; (M9) \citet{Reffert2015}; (M10) \citet{Gaspar2016}; (M11) \citet{Boeche2018}; (M12) \citet{Luck2007}; (M13) \citet{Luo2016} \\
 V mag, ST, Parallax are taken from SIMBAD	

\end{table*}

\begin{table*}
\centering
\contcaption{}
	\begin{tabular}{lccccccccccr} 
    \hline
 Stars & V & ST & $T_{eff}$ & log $g$ & [Fe/H] & Parallax & Date Of & Exposure & SNR  & Sky & Ref \\
 Names & mag &  & (K) & (cm/$s^2$) &  (dex) &  (mas) & Observation & Time (s) & & Conditions &   \\
\hline
HD17895 & 7.16 & M5III & 3294 & -0.021 & -- & 2.72 & 2016-12-20 & 2*(5*7) & 120 & clear sky & T1 \\
HD22689 & 7.16 & M5III & 3144 & -0.135 & -- & 3.85 & 2015-12-18 & 2*(3*4) & 77 & clear sky & T1 \\
HD26234 & 8.90 & M5III & 3191 & 0.390 & -- & 2.95 & 2015-01-13 & 2*(5*50) & 94 & clear sky & T1 \\
HD39983 & 8.26 & M5III & 3145 & 0.430 & -0.23 & 4.73 & 2016-12-19 & 2*(5*10) & 154 & clear sky & T1,M3 \\
HD46421 & 8.21 & M5III & 3225 & 0.215 & -- & 3.57 & 2016-12-19 & 2*(5*10) & 114 & clear sky & T1 \\
HD66175 & 7.04 & M5III & 3156 & 0.158 & -- & 3.41 & 2015-01-17 & 2*(5*12) & 128 & clear sky & T1 \\
HD103681 & 6.20 & M5III & 3215 & -0.056 & -- & 2.81 & 2016-12-19 & 2*(5*50) & 124 & clear sky & T1 \\
HD105266 & 7.18 & M5III & 3246 & 0.001 & -- & 3.69 & 2015-01-14 & 2*(5*10) & 83 & clear sky & T1 \\
HD64657 & 6.85 & M5/M6III & 3269 & 0.031 & -- & 3.78 & 2016-12-19 & 2*(5*6) & 114 & clear sky & T1 \\
HD65183 & 6.40 & M5/M6III & 3359 & 0.020 & -- & 2.93 & 2016-12-19 & 2*(5*6) & 94 & thin cloud & T1 \\
HD223608 & 8.86 & M5/M6III & 3228 & -- & -- & 0.57 & 2015-12-19 & 2*(3*15) & 136 & clear sky & T2 \\
HD7861 & 8.54 & M6III & 3259 & -- & -- & 4.20 & 2015-01-13 & 2*(5*40) & 100 & clear sky & T2 \\
HD18191 & 5.93 & M6III & 3336 & 0.332 & -0.24 & 9.28 & 2015-12-18 & 2*(3*4) & 98 & clear sky & T1,M5 \\
HD27957 & 8.03 & M6III & 3383 & 0.298 & -- & 2.38 & 2016-12-20 & 2*(5*40) & 127 & clear sky & T1 \\
HD70421 & 8.55 & M6III & 3120 & -0.112 & -- & 2.22 & 2016-12-19 & 2*(5*12) & 104 & clear sky & T1 \\
HD73844 & 6.67 & M6III & 3206 & 0.109 & -- & 6.40 & 2015-12-18 & 2*(3*5) & 87 & clear sky & T1 \\
HIP44601 & 9.20 & M6III & 3200 & -- & -- & 1.25 & 2015-03-02 & 2*(5*50) & 151 & clear sky & T2 \\
HIC55173 & 7.42 & M6III & 3288 & 0.280 & -- & 4.48 & 2016-12-19 & 2*(5*10) & 134 & clear sky & T1 \\
HIP57504 & 8.74 & M6III & 2920 & 0.011 & -- & 3.86 & 2016-12-19 & 2*(5*10) & 150 & clear sky & T1 \\
HD115322 & 7.21 & M6III & 3458 & -- & -- & 1.3 & 2015-01-16 & 2*(5*35) & 97 & clear sky & T2 \\
HD203378 & 7.32 & M6III & 3284 & 0.035 & -- & 3.23 & 2015-08-11 & 2*(3*20) & 57 & clear sky & T1 \\
HD43635 & 7.93 & M7III & 3240 & -- & -- & -- & 2015-12-19 & 2*(3*10) & 115 & clear sky & T3 \\
HIC51353 & 9.84 & M7III & 3224 & -- & -- & -0.39 & 2015-01-14 & 2*(6*70) & 65 & clear sky & T2 \\
HIC68357 & 9.03 & M7III & 3138 & 0.362 & -- & 3.58 & 2015-01-14 & 2*(6*60) & 91 & clear sky & T1 \\
HD141265 & 10.45 & M8III & 2701 & -- & -- & 2.35 & 2015-12-18 & 2*(3*20) & 82 & clear sky & T2 \\
 &  & & & & & & & & & & \\
\bf{SpeX} : &  & & & & & & & & & & \\
 &  & & & & & & & & & & \\

HD100006 & 5.54 & K0III & 4714 & 2.288 & -0.12 & 10.36 & -- & -- & 260 & -- & T1,M12 \\ 
HD9852 & 7.92 & K0.5III & 4750 & -- & -- & 1.58 & -- & -- & 210 & -- & T4 \\ 
HD25975 & 6.09 & K1III & 5022 & 3.320 & -0.20 & 22.68 & -- & -- & 244 & -- & T1,M4 \\ 
HD36134 & 5.78 & K1-III & 4519 & 1.893 & -- & 6.62 & -- & -- & 244 & -- & T1 \\ 
HD91810 & 6.53 & K1-III & 4561 & 2.059 & -- & 5.22 & -- & -- & 192 & -- & T1 \\ 
HD124897 & -0.05 & K1.5III & 4280 & 1.70 & -0.52 & 88.53 & -- & -- & 261 & -- & T5,M1 \\ 
HD137759 & 3.29 & K2III & 4570 & 2.248 & 0.03 & 32.23 & -- & -- & 215 & -- & T1,M1 \\ 
HD132935 & 6.69 & K2III & 4220 & 1.483 & -- & 3.59 & -- & -- & 248 & -- & T1 \\ 
HD2901 & 6.92 & K2III & 4319 & 1.850 & -0.02 & 4.32 & -- & -- & 278 & -- & T1,M3 \\ 
HD221246 & 6.15 & K3III & 4145 & 1.255 & -- & 3.61 & -- & -- & 184 & -- & T1 \\ 
HD178208 & 6.43 & K3III & 4315 & 1.803 & -- & 5.18 & -- & -- & 167 & -- & T1 \\ 
HD35620 & 5.05 & K3III & 4239 & 1.449 & 0.11 & 7.20 & -- & -- & 174 & -- & T1,M8 \\ 
HD99998 & 4.77 & K3+III & 3976 & 0.864 & -0.39 & 5.40 & -- & -- & 189 & -- & T1,M8 \\ 
HD114960 & 6.53 & K3.5III & 4130 & 1.795 & -- & 6.19 & -- & -- & 168 & -- & T1 \\ 
HD207991 & 6.85 & K4-III & 3837 & 1.002 & -- & 2.97 & -- & -- & 197 & -- & T1 \\ 
HD181596 & 7.51 & K5III & 3893 & 0.588 & -- & 1.24 & -- & -- & 168 & -- & T1 \\ 
HD120477 & 4.07 & K5.5III & 3962 & 1.263 & -0.23 & 12.38 & -- & -- & 167 & -- & T1,M8 \\ 
HD3346 & 5.12 & K6III & 3820 & 0.782 & -- & 5.29 & -- & -- & 166 & -- & T1 \\ 
HD194193 & 5.93 & K7III & 3819 & 0.850 & -- & 3.86 & -- & -- & 169 & -- & T1 \\ 
HD213893 & 6.73 & M0III & 3855 & 1.278 & -0.09 & 4.22 & -- & -- & 182 & -- & T1,M11 \\ 
HD204724 & 4.50 & M1+III & 3847 & 0.967 & -- & 8.28 & -- & -- & 176 & -- & T1 \\ 
HD120052 & 5.43 & M2III & 3598 & 0.488 & -- & 4.82 & -- & -- & 160 & -- & T1 \\ 
HD219734 & 4.86 & M2.5III & 3677 & 0.525 & 0.04 & 5.79 & -- & -- & 145 & -- & T1,M5 \\ 
HD39045 & 6.26 & M3III & 3582 & 0.871 & -- & 5.44 & -- & -- & 143 & -- & T1 \\ 
HD28487 & 6.99 & M3.5III & 3441 & 0.533 & -- & 3.59 & -- & -- & 132 & -- & T1 \\ 
HD4408 & 5.38 & M4III & 3492 & 0.155 & -- & 4.20 & -- & -- & 128 & -- & T1 \\ 
HD204585 & 5.95 & M4.5III & 3379 & 0.161 & -- & 4.90 & -- & -- & 117 & -- & T1 \\ 
HD27598 & 7.04 & M4III & 3490 & 0.489 & -- & 2.94 & -- & -- & 135 & -- & T1 \\ 
HD19058 & 3.39 & M4+III & 3479 & 0.302 & -0.15 & 10.60 & -- & -- & 124 & -- & T1,M4 \\ 
HD214665 & 5.16 & M4+III & 3476 & 0.482 & -- & 7.23 & -- & -- & 122 & -- & T1 \\ 
HD175865 & 4.00 & M5III & 3363 & 0.092 & 0.14 & 10.94 & -- & -- & 112 & -- & T1,M4 \\ 
HD94705 & 5.78 & M5.5III & 3371 & 0.379 & -- & 8.39 & -- & -- & 113 & -- & T1 \\ 
HD196610 & 5.89 & M6III & 3227 & 0.180 & -- & 8.56 & -- & -- & 114 & -- & T1 \\ 
HD108849 & 7.28 & M7-III & 2936  & -- & -0.34 & 5.53 & -- & -- & 88 & -- & T1,M13 \\ 
BRI2339-0447 & - & M7-8III & 3200 & -- & &-- & -- & -- & 76 & -- & T4 \\   

\hline
	\end{tabular}
\end{table*}

\subsection{Observations}
\begin{figure*}
	\includegraphics[scale=1.0]{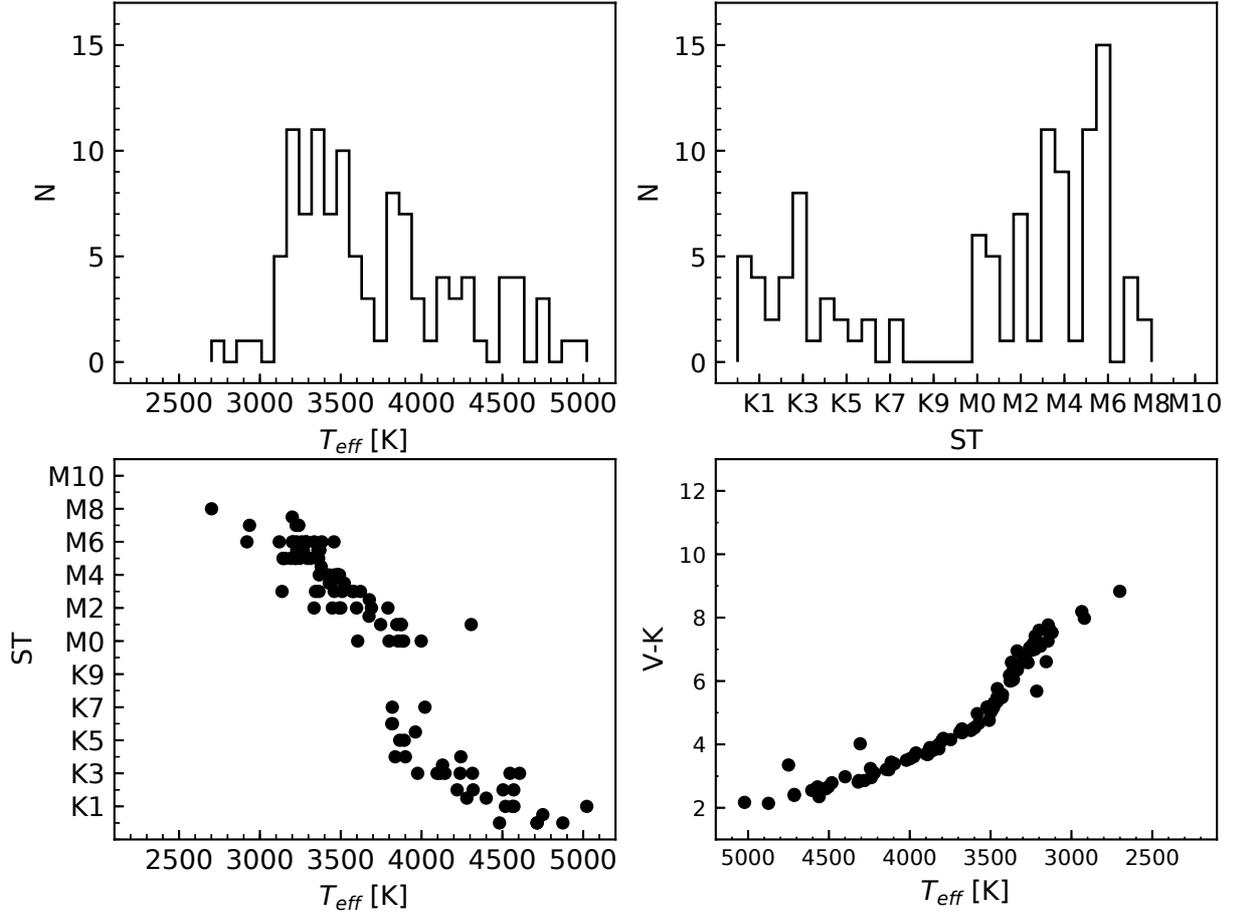}
    \caption{Effective temperature ($T_{eff}$) and Spectral type (ST) distribution of the sample are shown in the top-left and top-right panel respectively. $T_{eff}$ vs. ST and $T_{eff}$ vs. V$-$K for the sample are shown in the bottom-left and bottom-right panel respectively.}
    \label{fig1:sample_selection}
\end{figure*}

\begin{figure*}
	\includegraphics[scale=0.75]{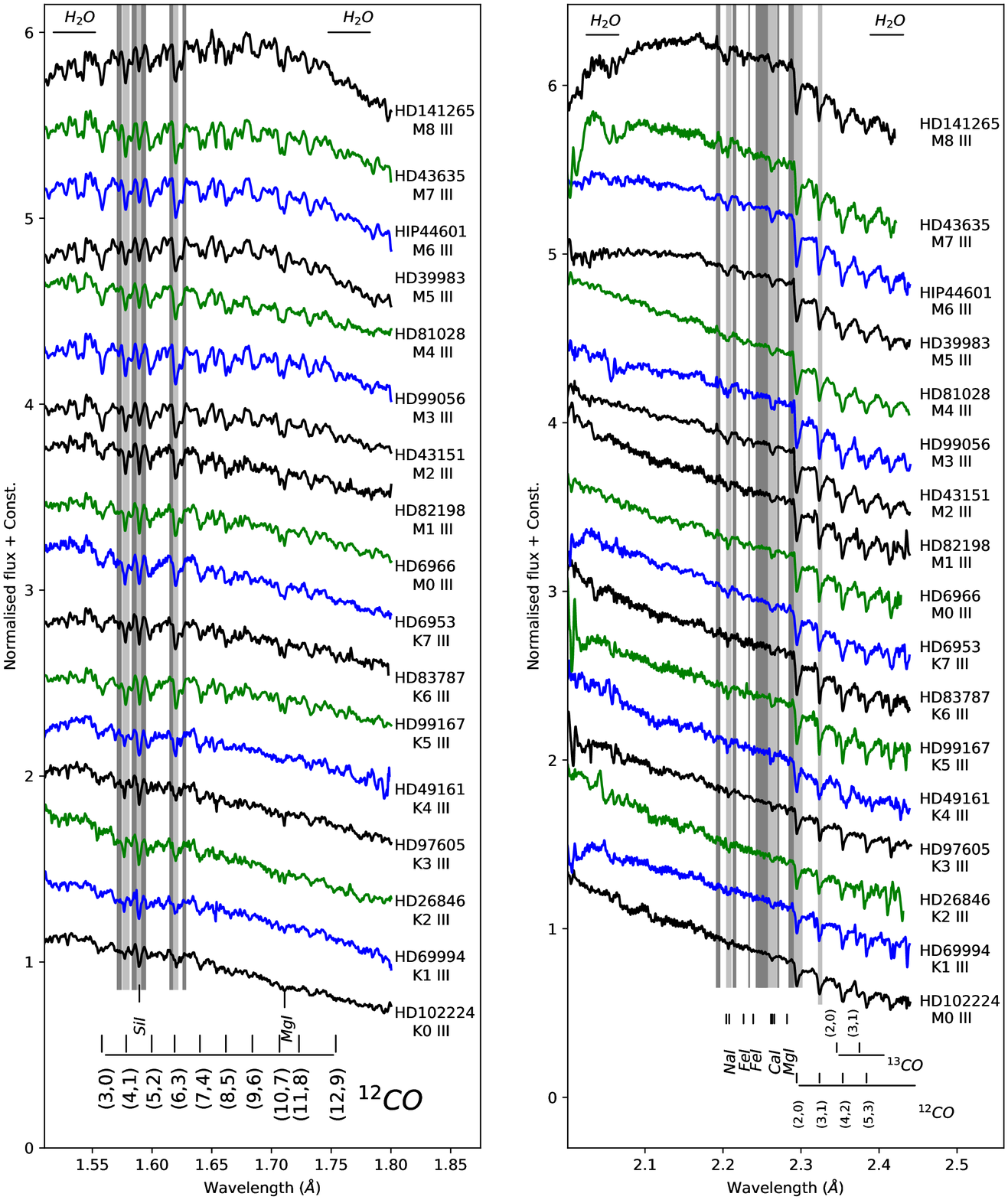}
    \caption{Subset of HK-band spectra of giants (K0$-$M8) observed with TIRSPEC instrument are shown in Figure. All the spectra have normalised to unity at 1.65 $\micron$ (H-band) and 2.17 $\micron$ (K-band), and offset by constant value with respect to the bottom-most spectrum for displaying purposes. The names of the stars and spectral types have been mentioned right end of the corresponding spectra. All the prominent features in HK-band are marked. The grey regions represent the continuum bandpasses, and the silver region represents the feature bandpasses as mentioned in this paper (see Table~\ref{tab:bandpass}).}
    \label{giants_hkspectra}
\end{figure*}

NIR spectra of seventy-two K$-$M giants are obtained using medium resolution TIFR Near-Infrared Spectrometer and Imager (TIRSPEC)  on the 2.0-m Himalayan Chandra Telescope (HCT) located at Hanle, India. Additional details of the TIRSPEC instrument can be found in \citet{Ninan2014}. The spectra are taken with cross-disperser mode (1.50$-$1.84 $\mu$m and 1.95$-$2.45 $\mu$m) with a slit width 1.97$\arcsec$ during several observing runs spanning over 2014 to 2017, and the log of observations is mentioned in Table~\ref{Table: Identification of Stars}. The spectra are taken at two different positions along the slit one after another, immediately to subtract the sky, and several frames are observed to improve the signal to noise ratio (SNR). The integration time varied from 4s to 100s depending on the magnitude of stars. The spectral type (ST) of sample stars spans from K0 to M8 with declination higher than $-$32 degrees. The main criterion is to populate the space with the sample stars of $T_{eff}$ from 2500 K to 5000 K. Special attention is given to observe the stars of M3 III or later.  About 60\% giants in our sample have ST M3 or beyond for better characterization in late M-region. We select very bright sample stars in order to get high SNR. Suitable standard stars of ST A0V to A1V are observed after each observation.

To populate our sample, we have used all available thirty-five giants spanning spectral range K0 to M8 in the IRTF spectral library \citep{Cushing2005, Rayner2009}. Those spectra are observed with the medium-resolution  ($\lambda/\Delta \lambda \sim$ 2000) SpeX infrared spectrograph in the wavelength range 0.8-2.4 $\mu$m mounted on the 3.0-m IRTF at Mauna Kea, Hawaii. Thus, we have assembled one hundred seven giants for the present study. We ignore the known Mira variables and OH/IR stars belonging to M-spectral types of the IRTF library from our study.

The photometric data of those stars are taken from literature as shown in Table~\ref{Table: Identification of Stars}. The T$_{eff}$ and log $g$ of the sample giants (97 out of 107) are uniformly taken from \citet{McDonald2017}, which are derived by comparing multi-wavelength archival photometry to BT-Settle model atmospheres. The uncertainties in their measurements are $\pm$125 K in T$_{eff}$.  The parameters of the rest 10 giants are taken from other literature as mentioned in Table~\ref{Table: Identification of Stars}. The metallicity of only 32 giants in our sample are available in the literature (see, Table~\ref{Table: Identification of Stars}). Figure~\ref{fig1:sample_selection} represents T$_{eff}$ and ST distribution of the sample, and their population in the T$_{eff}$ $-$ ST and T$_{eff}$ $-$ (V $-$ K) planes.

\subsection{Data Reduction} 
The spectroscopic analysis is done using APALL task of IRAF. The TIRSPEC data have been reduced with TIRSPEC pipe-line\footnote{https://github.com/indiajoe/TIRSPEC/wiki} \citep{Ninan2014}, and are cross-checked with the Image Reduction and Analysis Facility (IRAF \footnote{http://iraf.noao.edu/}). The data reduction consists of flat-fielding, sky subtraction, bad pixel correction, cosmic-ray removal, subtracting the pairs of images taken at two different slit positions, the wavelength calibration with Argon arc lamp, and finally, the spectrum extraction. To remove the telluric features of the Earth's atmosphere the spectra of program stars are divided by the spectra of the standard star, which is taken on the same night. Prior to division, all hydrogen lines are removed from the spectra of the standard stars by interpolating the stellar continua.  This is followed by the flux calibration of the target stars by using their Two Micron All Sky Survey (2MASS) H and K band photometric magnitudes.

\section{Equivalent widths measurement}

\begin{table*}
\centering
\caption{Definitions of Spectral Bands to Measure Equivalent Widths.}
\label{tab:bandpass}

\centering
\begin{tabular}{lcccr} 
   \hline
   \hline
Index & Feature & Feature  & Continuum  &  Ref. \\	 
      &         & Bandpass ($\mu$m) & Bandpass ($\mu$m) &  \\
\hline		

 SiI & Si~I (1.59 $\mu$m) & 1.5870-1.5910 & 1.5830-1.5870, 1.5910-1.5950 & 1 \\
CO1 & $^{12}$CO(2-0) (1.58 $\mu$m) & 1.5752-1.5812 & 1.5705-1.5745, 1.5830-1.5870 & 2 \\ 
CO2 & $^{12}$CO(2-0) (1.62 $\mu$m) & 1.6175-1.6220 & 1.6145-1.6175, 1.6255-1.6285 & 1 \\
NaI & Na~I (2.21 $\mu$m)  & 2.2040-2.2107 & 2.1910-2.1966, 2.2125-2.2170 & 3 \\
CaI & Ca~I (2.26 $\mu$m) & 2.2577-2.2692 & 2.2450-2.2560, 2.2700-2.2720 & 3 \\
CO3 & $^{12}$CO(2-0) (2.29 $\mu$m) & 2.2910-2.3020 & 2.2420-2.2580, 2.2840-2.2910 & 2 \\
CO4 & $^{12}$CO(3-1) (2.32 $\mu$m) & 2.3218-2.3272 & 2.2325-2.2345, 2.2695-2.2715 & 2 \\

\hline
\end{tabular}
\\
Ref : (1) \citet{Origlia1993}; (2) \it{This work}; (3) \citet{Frogel2001}  
\end{table*}

The standard definition of equivalent width (EWs) is
\begin{equation}
EW_\lambda = \int_{\lambda1}^{\lambda2}\bigg[1 - \frac{F(\lambda)}{F_c(\lambda)}\bigg]d\lambda
\end{equation}
Here, F($\lambda$) represents the flux density inside the feature bandpass from $\lambda1$ to $\lambda2$, $F_c(\lambda)$ represents the value of the local continuum \citep{Cesetti2013}. 
To measure EWs feature band and continua bands are adopted as shown in Table~\ref{tab:bandpass} and in Figure~\ref{giants_hkspectra}. Bands of $^{12}$CO at 1.58 $\mu$m are newly defined in this study. We compute the $^{12}$CO at 1.62 $\mu$m band-strength according to the recipe of \citet{Origlia1993}. Instead of four continua adopted by \citet{Frogel2001}  to measure $^{12}$CO at 2.29 $\mu$m absorption depth, we use here two continua as mentioned in Table~\ref{tab:bandpass}.  We adopt the feature bandpass of $^{12}$CO at 2.32 $\mu$m from \citet{Kleinmann1986}, however, we use two different continua bands instead of one continuum used in \citet{Kleinmann1986}. Different continuum bands are also verified as mentioned in \citet{Ivanov2004, Cesetti2013}, but better results are obtained from our selected band-passes. 

 Before computing EWs, the spectral features are corrected for the zero velocity by shifting. The EWs are estimated with the IDL script\footnote{\url{https://github.com/ernewton/nirew}} \citep{Newton2014}. In the script, pseudo-continuum, i.e. continuum in featured bandpass, is defined by fitting a straight line through the continuum bandpass and EWs are measured by numerically integrating (trapezoidal method) the flux within the feature bandpass. 
 
The average spectral resolution of TIRSPEC data is, R $\sim$ 1200. The spectral resolution varies with wavelengths and such resolution variation in TIRSPEC can be found in \citet{Ninan2014}. The resolution of SpeX data is R $\sim$ 2000. The SpeX spectra are degraded to the same spectral resolution as of TIRSPEC before all the indices are estimated. Uncertainties on the EWs are computed with the Monte Carlo approach in the IDL script. The script adds normally-distributed (Gaussian) random noise to the stars' spectrum by using RANDOMN function \citep{Newton2014}. Provided errors (photon, residual sky, and read noise) in SpeX pipeline are used for Gaussian random noise simulation to estimate uncertainties in EWs. But, our TIRSPEC pipeline does not provide such errors, and the errors are estimated using the technique provided by \citet{Stoehr2008}. The computed EWs of our sample are listed in Table~\ref{tab:all_measured_ews} along with their uncertainties. 

\section{RESULT AND DISCUSSION}
\subsection{Behaviour of spectral features} 

\begin{figure*}
	\centering
	\includegraphics[scale=0.70]{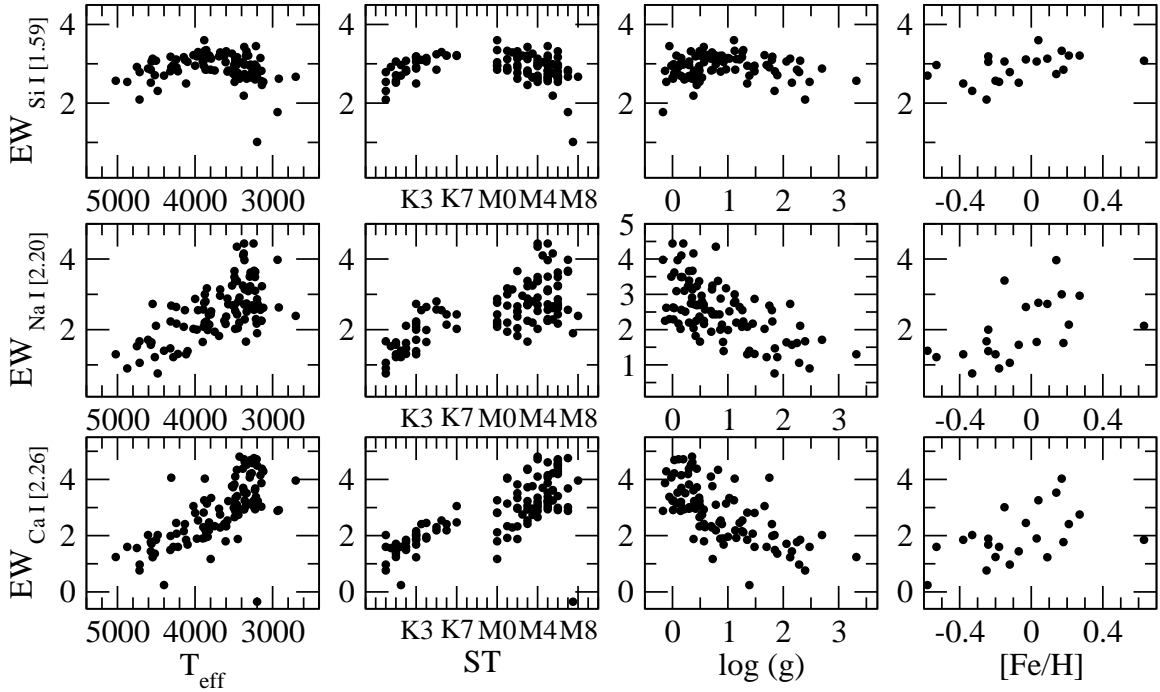}
    \caption{Behaviour of measured equivalent widths of atomic features with T$_{eff}$, ST, log $g$, and [Fe/H] are shown in this figure.}
    \label{Fig:behaviour_atomic_features}
\end{figure*}

\begin{figure*}
	\includegraphics[scale=0.70]{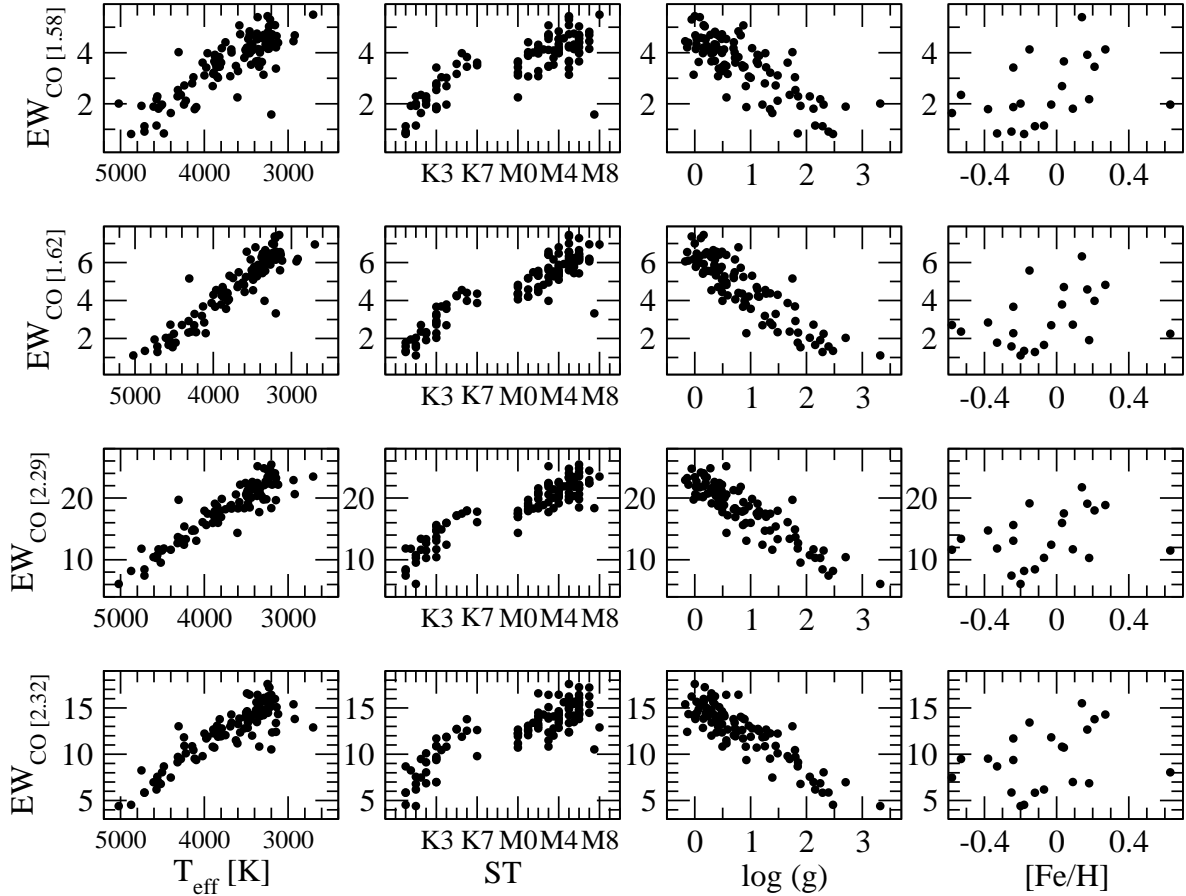}
    \caption{Behaviour of measured equivalent widths of molecular features with T$_{eff}$, ST, log $g$, and [Fe/H] are shown.}
    \label{Fig:behaviour_molecular_features}
\end{figure*}

\begin{figure*}
	\includegraphics[scale=0.67]{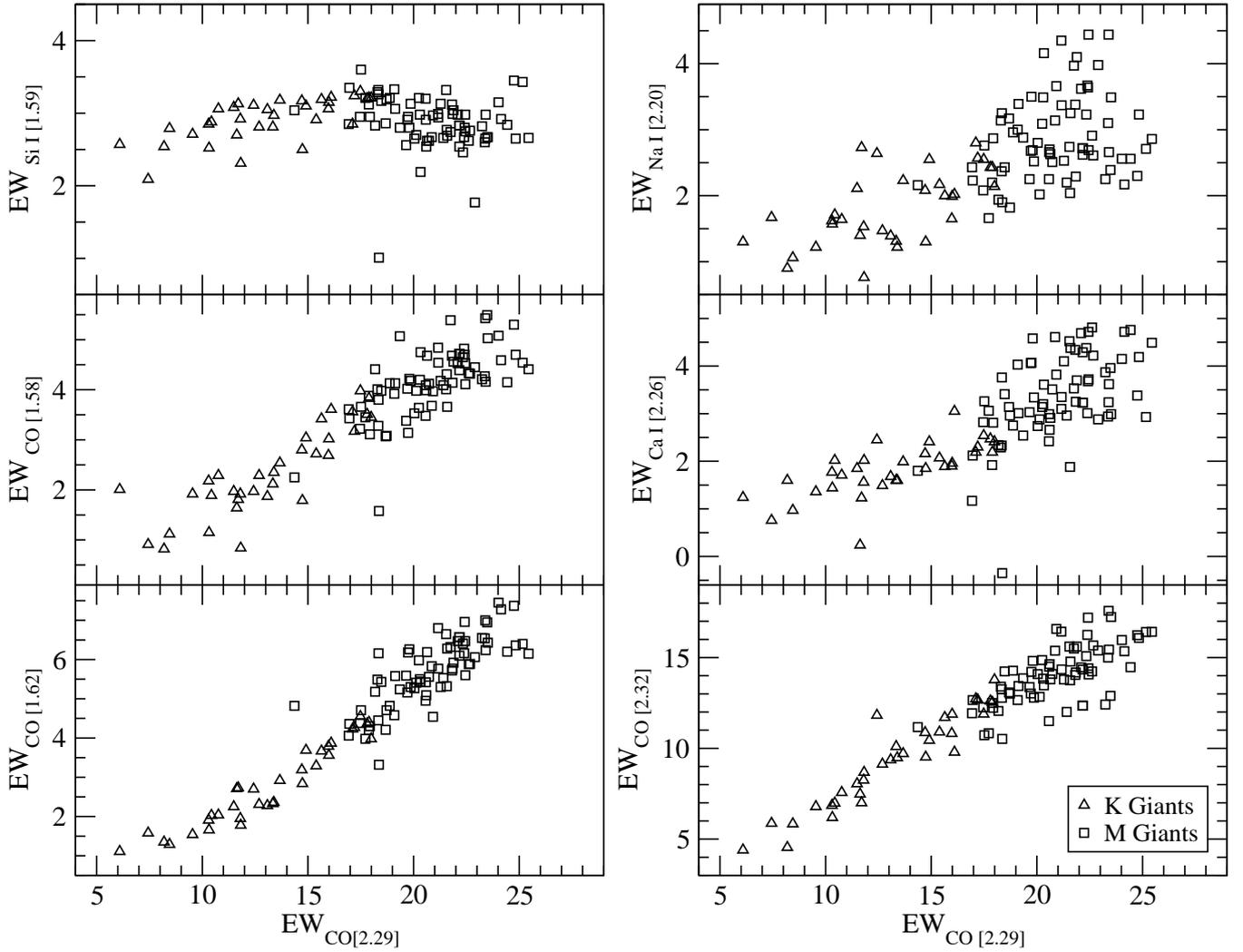}
    \caption{Measured equivalent widths of $^{12}$CO at 2.29 $\mu$m feature versus measured equivalent widths of Si I, Na I, Ca I and $^{12}$CO at 1.58, 1.62 and 2.32 $\mu$m.}
    \label{Fig:index-index}
\end{figure*}

We have studied here the behaviour of spectral signatures of the giants with stellar atmospheric parameters like T$_{eff}$, ST, log $g$ and [Fe/H].
The most prominent atomic lines of HK band spectra are Si I at 1.59 $\mu$m, Na I at 2.20 $\mu$m, and Ca I at 2.26 $\mu$m as shown in Figure~\ref{giants_hkspectra}. EWs of those lines are estimated using the methods as described in section~3. The behaviour of those lines with T$_{eff}$, ST, log $g$, and [Fe/H] are shown in Figure~\ref{Fig:behaviour_atomic_features}. The Si I lines is one of the strongest absorption features in K giants, and it's strength steadily increases as the temperature decreases from 5000 to 4000 K, after that remains unchanged up to 3500 K and decreases further below 3500 K. The corresponding behaviour of Si I with ST is also observed,  e.g., increasing in the range K0$-$K7, unchanged to M4 and decreasing further, and it appears insensitive to the log $g$.

The strengths of Na I and Ca I strongly depend on $T_{eff}$ and show an increasing trend with decreasing T$_{eff}$ as found by others \citep{Kleinmann1986, Ramirez1997, Forster2000, Frogel2001, Ivanov2004, Rayner2009, Cesetti2013}. Correlation of Na I line with ST shows increasing strength for K-giants, but no trend is conclusive for M giants. In the case of Ca I, the strength indicates an increasing trend with ST, similar to T$_{eff}$. The Na I and Ca I lines get stronger with decreasing log $g$.

There is a significant dispersion in both correlations (T$_{eff}$ and ST) for all the atomic lines. The poor band strengths in our medium-resolution spectra have some important role in such dispersion. Furthermore, contamination from other atomic lines are also affecting such studies, and relatively better spectral resolution are required for better characterization. \citet{Origlia1993} mentioned that Si I feature is somewhat contaminated by OH line at lower temperature and strength of OH line dominated beyond M2, i.e., T$_{eff}$ $\leq$ 3800 K. The Na doublet at 2.2 $\mu$m are blended with metallic lines like Si I (2.2069 $\mu$m), Sc I (2.2058 and 2.2071 $\mu$m) and V I (2.2097 $\mu$m) in our medium-resolution spectra, and such dispersion in M-giants might be related to other lines behaviour \citep{Wallace1996}. For late M-giants, few low excitation lines like Ti I (2.2627 and 2.2639 $\mu$m) and Sc I (2.2642 and 2.2663 $\mu$m) contaminate the Ca triplet at 2.26 $\mu$m.  

In the 1.5$-$2.4 $\mu$m regions, the first-overtone ($\Delta\nu$=2) and the second-overtone ($\Delta\nu$=3) band heads of $^{12}$CO  are the dominant features in K$-$M giants, and show increasing strength from K to M. In Figure~\ref{Fig:behaviour_molecular_features}, comparative behaviour of different CO bandheads (1.58 (CO1), 1.62 (CO2), 2.29 (CO3) and 2.32 (CO4) $\mu$m)  with  T$_{eff}$, ST, log $g$ and [Fe/H] show an increasing trend of band strengths with decreasing T$_{eff}$, early to late ST, and decreasing value of log $g$ (see, e.g., \citealt{Origlia1993, Ramirez1997, Cesetti2013}).  The behaviour of EWs with metallicity [Fe/H] in Figure~\ref{Fig:behaviour_atomic_features} and Figure~\ref{Fig:behaviour_molecular_features} does not show any conclusive trend as expected because most of our sample belong to solar-neighbourhood giants.

To investigate the origin of the dispersion especially in Figure~\ref{Fig:behaviour_atomic_features}, we plot index-index relations as shown in Figure~\ref{Fig:index-index}. Figure~\ref{Fig:index-index} shows a tight index-index correlation at least for CO[1.62] $-$ CO[2.29] and CO[2.32] $-$ CO[2.29] as CO-bands are strong features in the medium-resolution spectra. Small dispersion of CO index-index correlations might be due to various reasons, such as the variation of abundance ratios, residuals of telluric lines. Note that we discard the known Mira variables and OH/IR stars belonging to M-spectral type of the IRTF library due to their large variability, and they might have different behaviour compared to the static giants \citep{Lancon2000}.

\subsection{Empirical Calibrations}
\begin{table*}
\caption{Comparison between Goodness of Fit for various correlations.}
\label{tab:Goodness of fit}
\resizebox{0.80\textwidth}{!}{
\begin{tabular}{lccccccccc} 
   \hline
   \hline
Index & T & N & R & Rsqr & SEE & a0$\dagger$ & a1$\dagger$ & a2$\dagger$ & Remarks* \\	
\hline
 & & & & & T$_{eff}$ & = f(EW) :  & & &\\
\hline
$^{12}$CO (4-1)   & 107 & 101 & 0.90 & 0.82 & 207 & 5114 $\pm$ 70 & -390 $\pm$ 19 & - & 1 \\
1.58 $\mu$m       &  98 &  93 & 0.90 & 0.82 & 197 & 5070 $\pm$ 69 & -372 $\pm$ 19 & - & 2 \\
(CO1)             &  70 &  67 & 0.90 & 0.82 & 177 & 5039 $\pm$ 68 & -346 $\pm$ 20 & - & 3 \\
\hline	
$^{12}$CO (6-3)   & 107 & 102 & 0.96 & 0.92 & 140 & 5049 $\pm$ 42 & -279 $\pm$ 8 & - & 1 \\
1.62 $\mu$m       &  98 &  95 & 0.96 & 0.92 & 130 & 5038 $\pm$ 40 & -274 $\pm$ 8 & - & 2 \\
(CO2)             &  70 &  67 & 0.96 & 0.92 & 124 & 5092 $\pm$ 45 & -287 $\pm$11 & - & 3 \\
\hline	 
$^{12}$CO (2-0)   & 107 & 100 & 0.96 & 0.93 & 130 & 5619 $\pm$ 54 & -103 $\pm$ 3 & - & 1 \\
2.29 $\mu$m       &  98 &  93 & 0.97 & 0.93 & 119 & 5563 $\pm$ 51 &  -99 $\pm$ 3 & - & 2 \\
(CO3)             &  70 &  67 & 0.97 & 0.94 & 104 & 5571 $\pm$ 53 &  -99 $\pm$ 3 & - & 3 \\
\hline
$^{12}$CO (3-1)   & 107 & 100 &0.94 &0.88 &166 &5603 $\pm$ 71 & -149 $\pm$ 5 & - &1 \\
2.32 $\mu$m       &  98 &  94 &0.95 &0.90 &147 &5549 $\pm$ 64 & -143 $\pm$ 5 & - &2 \\
 (CO4)            &  70 & 65  &0.95 &0.91 &127 &5532 $\pm$ 65 & -140 $\pm$ 6 & -  &3 \\            
\hline
 & & & & & log $g$ & = f(EW) :  & & & \\
\hline
CO2 & 97 & 92 & 0.91 & 0.82 & 0.29 & 2.69 $\pm$ 0.01 & -0.40 $\pm$ 0.02 & - & 1\\
\hline
CO3 & 97 & 93 & 0.93 & 0.86 & 0.29 & 3.75 $\pm$ 0.13 & -0.16 $\pm$ 0.01 & - & 1\\
\hline
 & & & & & T$_{eff}$ & = f(EW, log $g$) :   & & & \\
 \hline
CO2 & 97 & 90 & 0.99 & 0.97 & 78 & 4148 $\pm$ 72 & -142 $\pm$ 11 & 315 $\pm$ 25 & 1 \\
\hline
CO3 & 97 & 92 & 0.98 & 0.96 & 90 & 4465 $\pm$ 116 & -54 $\pm$ 5 & 308 $\pm$ 30 & 1 \\
\hline
\end{tabular}}

T - total nos. of data points; N - no. of points used for fitting after eliminating 2$\sigma$ outlayers \\ 
R - correlation coefficient; Rsqr - coefficient of determination;
SEE - standard error of estimate \\ 
$\dagger$ T$_{eff}$ = a0 + a1 $\times$ EWs + a2 $\times$ log $g$  \\
*1 - Fitting with all the sample stars \\
*2 - Fitting with sample stars; T$_{eff}$ $\geq$ 3200   \\
*3 - Fitting with sample stars;  T$_{eff}$ $\geq$ 3400  \\
\end{table*}

\subsubsection{Correlation between Effective Temperature and Equivalent Width}

The most strong  CO(2$-$0) bandhead at 2.29 $\mu$m has been widely used as a stellar $T_{eff}$ indicator. Several index definitions have been adopted to measure its strength (see \citealt{Kleinmann1986, Ramirez1997, Frogel2001, Blum2003, Maness2007, Marmol2008}). Different index definitions lead to overestimation of the stellar temperatures (see \citealt{Pfuhl2011}). \citet{Pfuhl2011} computed the CO strength according to the recipe of \citet{Frogel2001} to determine T$_{eff}$ using thirty-three giants with ST G0$-$M7, and have found smaller systematic error than other definitions. Other $^{12}$CO bandheads at 2.32 $\mu$m and 1.62  $\mu$m are also used as a reasonable good temperature indicator (see \citealt{Kleinmann1986, Origlia1993, Ivanov2004, Schultheis2016}). \citet{Schultheis2016} showed $^{12}$CO(3$-$1) bandhead at 2.32 $\mu$m is an excellent temperature indicator in alternative of the strong $^{12}$CO bandhead at 2.29 $\mu$m. 
  
\begin{figure*}
	\includegraphics[scale=0.55]{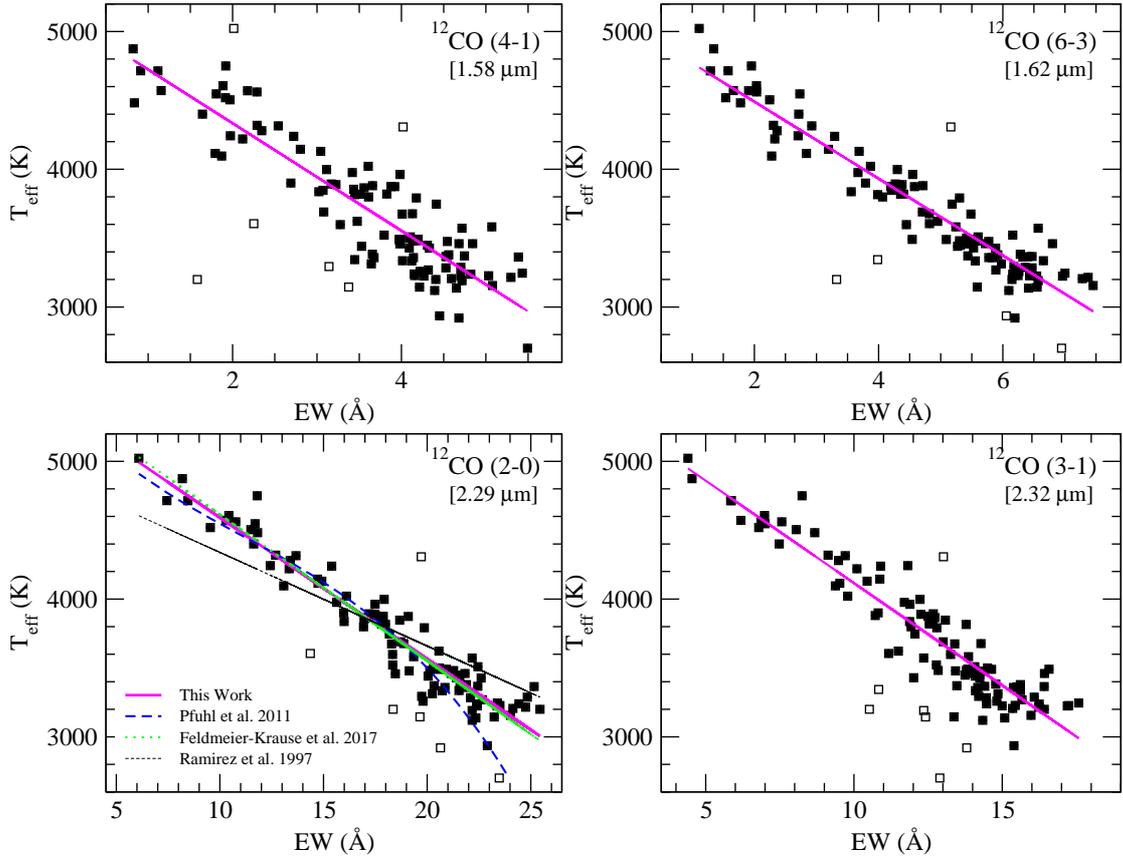}
    \caption{Figure shows the relation between T$_{eff}$ and EWs of the $^{12}$CO (4-1) at 1.58 $\mu$m, (6-3) at 1.62 $\mu$m, (2-0) at 2.29 $\mu$m and (3-1) at 2.32 $\mu$m.  For $^{12}$CO at 2.29 $\mu$m, we compare our results with literature. The square symbol represents all the stars of our sample. Black dot represents the stars from our sample used for empirical relation. The pink solid line shows our best fit relation.}
    \label{teff_vs_co}
\end{figure*}

We used all of the four bandheads CO1, CO2, CO3, and CO4 for new empirical relations of the giants, and for relative comparison of their effectiveness. Following \citet{Origlia1993} and \citet{Frogel2001}, we have used the bandpasses as mentioned in Table~\ref{tab:bandpass}. In case of CO1, we have defined here new bandpasses as in Table~\ref{tab:bandpass} that has not explored earlier. For CO3 and CO4, we have used two bands of the continuum from \citet{Frogel2001}, where the authors had used four bands of the continuum. The estimated EWs for all the sample stars are listed in Table~\ref{tab:all_measured_ews}. The EW of COs is plotted against T$_{eff}$ shown in Figure~\ref{teff_vs_co}. To establish the empirical relation between EW of COs and T$_{eff}$, a linear fit is explored for each bandhead separately using the linear equation T$_{eff}$ = a0 + a1 $\times$ EWs (where, a0, a1 are the coefficients of the fit). The 2$\sigma$ outliers are excluded for such fittings. Three different cases are excised for the best-fit, where case 1 is considered for all the 107 giants in our sample, case 2 for T$_{eff}$ $\geq$ 3200 with 98 giants and case 3 for T$_{eff}$ $\geq$ 3400 with 70 giants. The result of fitting in three different cases are listed in Table~\ref{tab:Goodness of fit}. The best-fit is judged by the three parameters $-$ correlation coefficient(R), the coefficient of determination (Rsqr) and the standard error of estimate (SEE). In case 1 (all the sample), SEE are 207 K, 140 K, 130 K, and 166 K for CO1, CO2, CO3, and CO4, respectively. In a comparison with all four bandheads, the SEE is minimum in case of strong bandhead CO3. We find that a better fit is obtained by narrowing down the temperature range and SEE improves from case 1 to case 3 for all bandheads. The least-square linear fits for case 1 only are shown in Figure~\ref{teff_vs_co}. For comparison, the existing relations in the literature are also over-plotted in Figure~\ref{teff_vs_co}, where the green dot line is the linear-fit from \citet{Feldmeier2017}, blue dash line is the three-degree polynomial fit of the \citet{Pfuhl2011} and black dot-dashed line is the linear fit from \citet{Ramirez1997}.

To establish the empirical relations, \citet{Feldmeier2017} used 69 stars with luminosity classes II-IV at a R $\sim$ 3310$-$4660, \citet{Pfuhl2011} used 33 giants at R $\sim$ 2000 and R $\sim$ 3000, and \citet{Ramirez1997} used 43 giants at R $\sim$ 1380 and R $\sim$ 4830. Our correlation with T$_{eff}$ $-$ CO3 differs significantly from the correlation of \citet{Ramirez1997}, and the difference could be due to different bandpass and continuum used to measure EWs. However, the correlation of T$_{eff}$ $-$ CO3 agrees well with that of \citet{Pfuhl2011} for T$_{eff}$ > 3000 K and \citet{Feldmeier2017}.  However, we reproduced almost the same or better correlation with lower residual scatter (SEE) in spite of using the lower resolution spectra. It is important to note here that spectral resolution (i.e. R $\sim$ 3310$-$4660 vs. R $\sim$ 1200) is insensitive to the T$_{eff}$ $-$ CO3 correlation as seen in \citet{Feldmeier2017}. Also, EWs of CO3 (2.29) are estimated using the two continuum bands, out of four continua as in \citet{Frogel2001}. However, our established correlation shows no significant variation compared with the relations of \citet{Pfuhl2011} and \citet{Feldmeier2017} as shown in Figure~\ref{teff_vs_co}, where the authors had used four continua of \citet{Frogel2001}. Thus, two continua could be used instead of four to calculate EWs of CO3 without any systematic offset. 

We also investigate simple parametrizations of the multi-line functions (e.g., CO1$-$SiI, CO1/SiI, CO2/CO1, CO2$-$CO1, CO3$-$(NaI+CaI), CO3/(NaI+CaI), etc.), and perform least-square regression to test the correlation with T$_{eff}$ for each combination qualitatively. No combining feature results significant improvement of the relationship discussed above. It is noted that the correlation between combined features follows the trends of the stronger feature of that combination. Further discussion on the combined features is therefore excluded.

\begin{figure*}
	\includegraphics[scale=0.60]{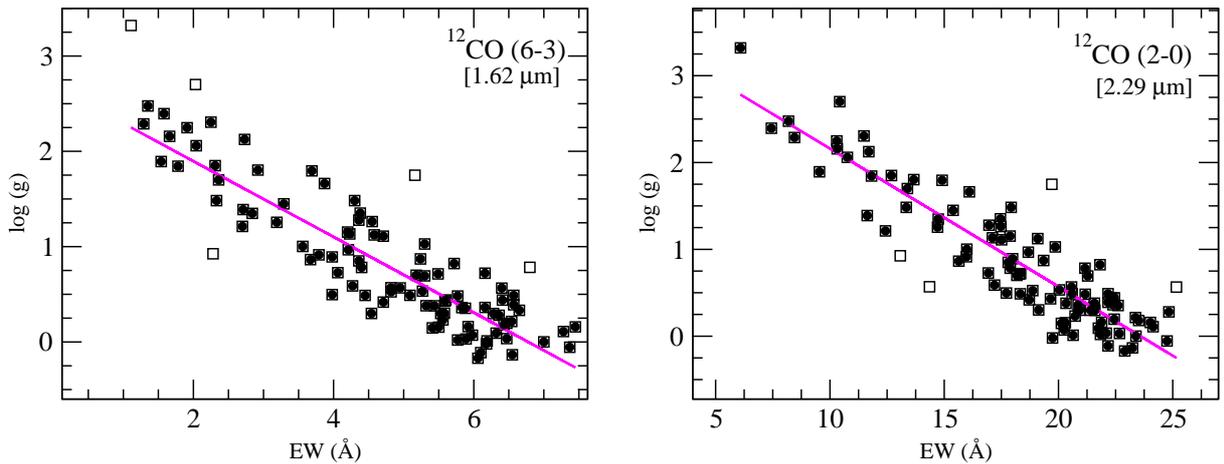}
    \caption{Figure shows correlation between log $g$ and EWs of the $^{12}$CO (6$-$3) at 1.62 $\mu$m and (2$-$0) at 2.29 $\mu$m. The square symbol represents all the stars of our sample. Black dot represents the stars from our sample used for empirical relation. The pink solid line shows our best fit relations.}
    \label{Fig:logg_metallicity}
\end{figure*}

\subsubsection{Correlation between Surface Gravity and Equivalent Width}

To establish the empirical relation between EWs and log $g$, a linear fit is explored for CO2 and CO3 bandhead separately using the linear equation log $g$ = $a_0$ + $a_1$ $\times$ EWs. We have used here CO2 and CO3 bands as their SNR are relatively better than other CO bands. Among 107 giants, 97 have known log $g$ in the literature and we have used here for the fit. We excluded the limiting 2$\sigma$ outliers for fitting. The least-square linear fits are shown in Figure~\ref{Fig:logg_metallicity}. The number of the stars used for the fit after 2$\sigma$ clipping and the coefficients of fit are listed in Table~\ref{tab:Goodness of fit} along with SEE. The best-fit is judged on the basis of SEE, which is 0.29 for both bands. Our study suggests that both CO2 and CO3 are good log $g$ indicator and the result differs from \citet{Origlia1993}, who demonstrated that CO2 is a better representative of log $g$ than CO3 from the behaviour of synthetic spectra. 

\subsubsection{Effect of Surface Gravity on Effective Temperature vs Equivalent Width correlation}
To take into account the effects of log $g$ on the calibration of T$_{eff}$ and the EWs of $^{12}$CO at 1.62 and 2.29 $\mu$m, we recalibrate the empirical relations as,
\begin{equation}
    T_{eff}=a_0 + a_1 {EW} + a_2 \log {g}.
    \label{eq:ew-temp-logg}
\end{equation}
where, $a_i$ (i=0,...,2) are the coefficients of the fit obtained iteratively. The SEE corresponds to 78 and 90 K for CO2 and CO3, respectively. The results of fitting are listed in Table~\ref{tab:Goodness of fit}. To test the effects of log $g$ quantitatively, we fit the equation~(\ref{eq:ew-temp-logg}) without considering log $g$ i.e. making $a_2$=0. The SEE is equivalent to 132 K for both cases. It is noted that the SEE is improved significantly when the effect of log $g$ is considered in the T$_{eff}$ vs EWs correlation. However, we did not consider the metallicity effect on T$_{eff}$ $-$ CO correlation since metallicity is unavailable for most of the stars in our sample. \citet{Schultheis2016} found no critical metallicity dependence on the T$_{eff}$ $-$ CO[2.29] correlation in the temperature range 3200$-$4500 K within metallicity range $-$1.2 to +0.5 dex.

\begin{figure*}
	\includegraphics[scale=0.60]{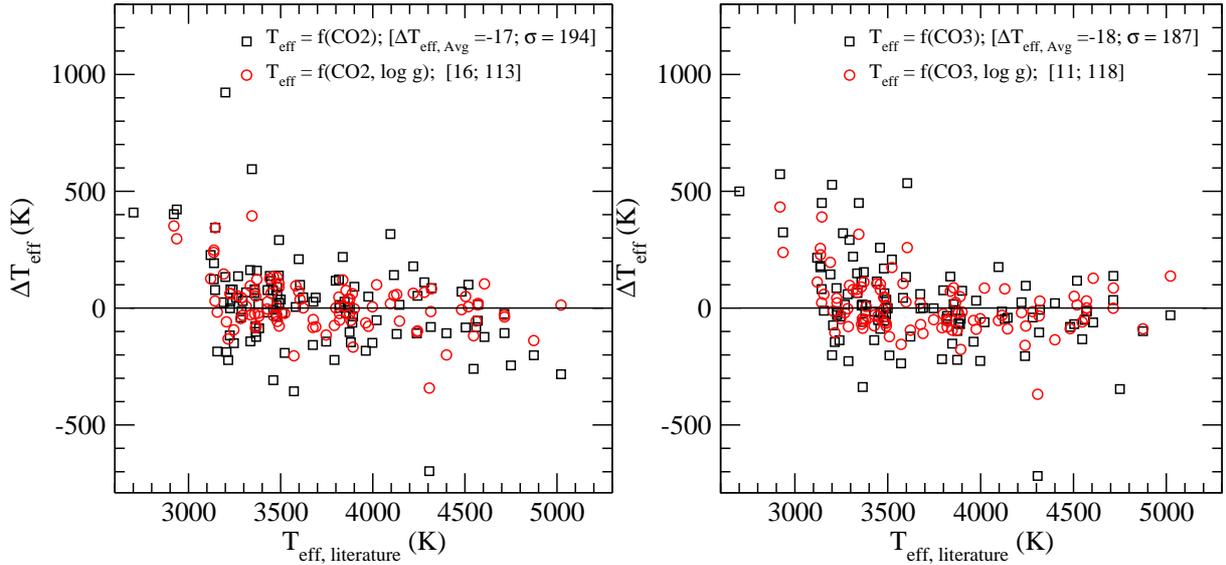}
    \caption{Comparison of T$_{eff}$ given in the Table~\ref{Table: Identification of Stars} and those derived from different established relations. We investigate the dependency of EWs and log $g$ on the derived T$_{eff}$. It shows that log $g$ significantly affects the results.}
    \label{Fig:all_correlation_comparison_for_teff}
\end{figure*}

\begin{figure*}
	\includegraphics[scale=0.60]{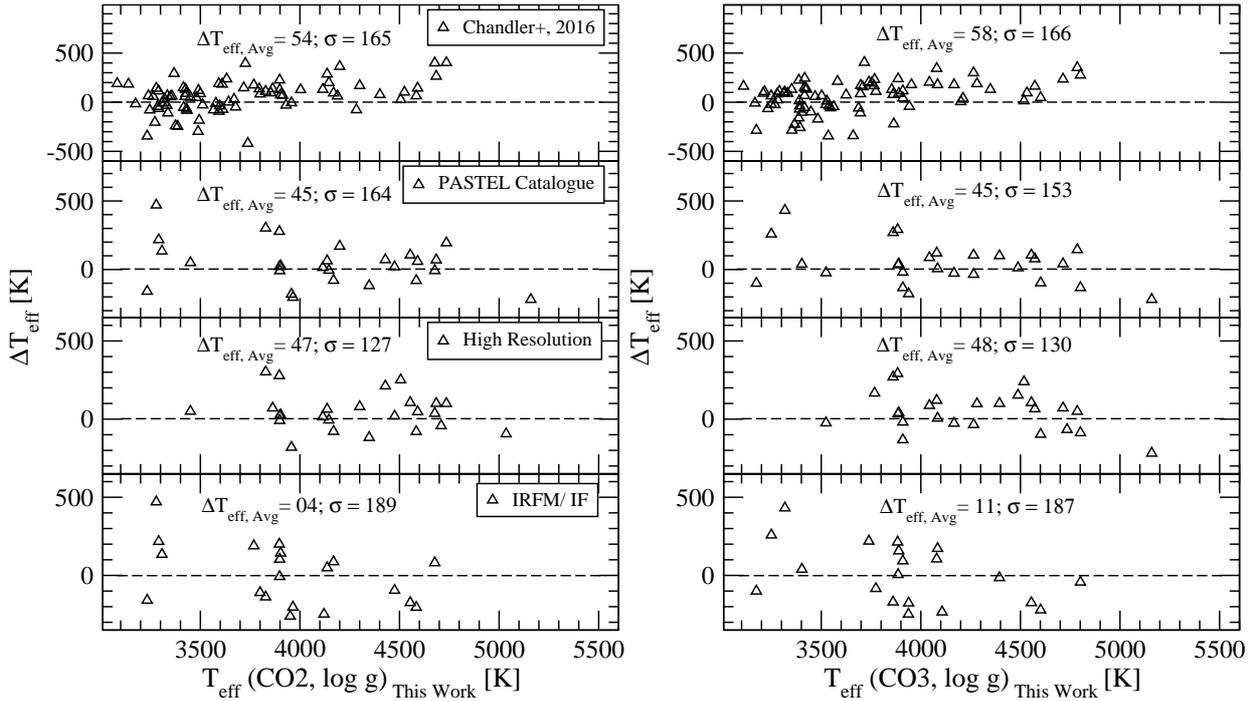}
    \caption{Comparison of T$_{eff}$ estimated from previous studies and those derived from our T$_{eff}$ $-$ CO $-$ logg relation. We provide the comparison for both CO2 (left panel) and CO3 (right panel). The residuals are plotted in Figure.}
    \label{Fig:Teff_comparison_from_different_literature}
\end{figure*}

\begin{figure*}
	\includegraphics[scale=0.60]{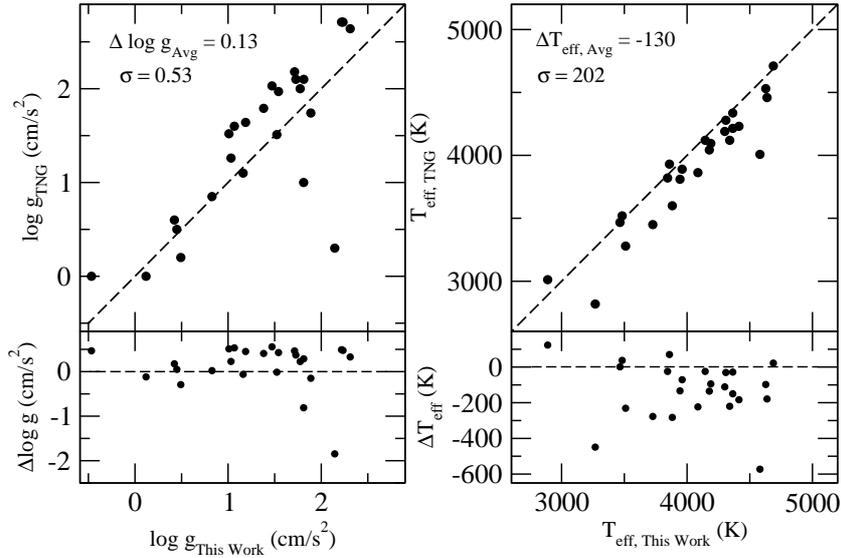}
    \caption{Comparison of [Fe/H], log $g$ and T$_{eff}$ derived by \citet{Marmol2008} and those derive from our calibrations using TNG spectra. The residuals are plotted at the bottom panel.}
    \label{Fig:comparison_from_TNG_spectra}
\end{figure*}

The T$_{eff}$ obtained from our empirical calibrations are compared with the previous published values of T$_{eff}$ estimated using various techniques. It is to be noted that we obtain the best T$_{eff}$ considering the effect of log $g$ along with the EWs of CO as described earlier. Hence, we derive values of T$_{eff}$ from the equation~\ref{eq:ew-temp-logg} using both CO2 and CO3 features, and compare distinctly with the literature values as shown in Figure~\ref{Fig:Teff_comparison_from_different_literature}. First, we focus on the Catalog of Earth-Like Exoplanet Survey Targets (CELESTA), a database of habitable zones around 37000 nearby stars \citep{Chandler2016}. We have 85 giants in common with our current sample of giants, but 5 of them (HD92620, HD115322, HD7861, HIP44601, HD141265) have not been considered owing to absence of log $g$ values in the literatures. We find that their T$_{eff}$ are on average $\sim$ 50 K cooler than our measurements, with a standard deviation, $\sigma$ $\sim$ 165 K for both (CO2 and CO3) cases. A total of 27 giants are found in common with PASTEL catalogue \citep{Soubiran2016}.  We find that the T$_{eff}$ of giants obtained in this work are on average 45 K warmer than the measurement in the PASTEL catalogue, with $\sigma$ $\sim$ 155 K. Estimation of T$_{eff}$ from high resolution spectra are available for 25 giants with our current sample\footnote{HD54810, HD137759 \citep{Jofre2015}; HD99283 \citep{Reffert2015}; HD102224, HD70272, HD60522, HD124897 \citep{Hekker2007}; HD69994, HD26846, HD97605, HD83787, HD91810, HD178208 \citep{Feuillet2016}; HD85503 \citep{Bruntt2011}; HD30834, HD92523, HD49161, HD99167, HD35620, HD99998, HD120477 \citep{McWilliam1990}; HD100006 \citep{Luck2007}; HD25975, HD19058 \citet{Smith1986}; HD207991 \citet{Kovtyukh2007}.}. A comparison of common objects  provide an average difference of 16 K (CO2) and 11 K (CO3), with $\sigma$ = 113 K (CO2) and 118 K (CO3). T$_{eff}$ of 4 giants are derived from IRFM method\footnote{HD54810 \citep{Blackwell1998}; HD219215, HD35620, HD99998 \citep{Alonso1999a}} and 16 giants are measured from interferometric data\footnote{HD102224, HD85503, HD92523, HD70272, HD99167, HD6953, HD38944, HD60522, HD216397, HD137759, HD120477, HD3346 \citep{Borde2002}, HD18191, HD175865, HD196610, HD108849 \citep{Dyck1998}}. The mean difference is $\sim$ 10 K, with $\sigma$ $\sim$ 180 K considering all the 20 giants.

\subsection{Application of our empirical relations}

To inspect reliability of our empirical relations, we estimate T$_{eff}$ and log $g$ from the spectra (R $\sim$ 1250) of K$-$M giants observed with Near Infrared Camera Spectrometer (NICS) on 3.58 m Telescopic Nazionale Galileo (TNG) at Roque de los Muchachos Observatory, La Palma, Spain \citep{Marmol2008}. A total of 25 K$-$M giants yield the opportunity to compare the parameters measured from our empirical relations with that of literature values\footnote{https://webs.ucm.es/info/Astrof/ellipt/CO.html}. The log $g$ is estimated from log $g$ $-$ CO3 relation. The results are in good agreement with average difference, $\Delta \log g_{Avg}$ = 0.13 $cm/s^2$ and standard deviation, $\sigma$ = 0.53 $cm/s^2$. The T$_{eff}$ are estimated using the measured log $g$ and CO3 from equation~\ref{eq:ew-temp-logg}. The T$_{eff}$ are on average 130 K cooler than literature value with a standard deviation, $\sigma$ = 202 K. Excluding the two giants, HD232708 (residual=572 K), a long period variable and HD126327 (residual=448 K), an asymptotic giant branch star, the $\Delta T_{eff, Avg}$ and $\sigma$ reduce to -97 K and 146 K respectively. The origin of this discrepancies might be due to the fact that pulsating long period variables behave differently than the static giants \citep{Bessell1989, Alvarez1998, Lancon2000, Ghosh2018}. The dispersion of two fundamental parameters from our measurements is shown in Figure~\ref{Fig:comparison_from_TNG_spectra}.

\section{Summary and Conclusions}

We have constructed a new medium resolution (R $\sim$ 1200) NIR (1.50$-$2.45 $\mu$m) spectral library of 72 K$-$M giant stars with the aim of populating existing NIR stellar libraries with cool giants specifically after the M3 spectral type. The EWs of prominent atomic (Si I at 1.59 $\mu$m, Na I at 2.20 $\mu$m, Ca I at 2.26 $\mu$m) and molecular ($^{12}$CO first overtone bandheads at 2.29 $\mu$m, 2.32 $\mu$m and, second overtone bandheads at 1.58 $\mu$m, 1.62$\mu$m) are estimated. We have studied here the behaviour of those EWs with the fundamental parameters (e.g., effective temperature, spectral type, surface gravity, and metallicity). The main results are summarized as

\begin{enumerate}

\item We obtained reliable new empirical relations between the EWs of $^{12}$CO bandheads and T$_{eff}$. We found that the $^{12}$CO first overtone band at 2.29 $\mu$m and second overtone band at 1.62 $\mu$m are reasonably good temperature indicator above 3400 K. This relation is also insensitive to the spectral resolution, and therefore, could be used more generally.

\item We present the empirical calibrations between the EWs of $^{12}$CO bandheads (CO2 and CO3)  and log $g$. Our study suggests that both $^{12}$CO are a very good indicator of log $g$. 

\item We find that the significant improvement of empirical relations between $^{12}$CO and T$_{eff}$ on the inclusion of log $g$, and more reliable T$_{eff}$ could be predicted. However, we do not investigate the metallicity effects of these correlations from such medium-resolution spectra in narrow metallicity range of our sample. Further investigation regarding metallicity from high-resolution spectra would be greatly appreciated.

\end{enumerate}

\section*{Acknowledgements}

The authors are very much thankful to the reviewer, Dr. R. Peletier, for his critical and valuable comments, which helped us to improve the paper. This research work is supported by S N Bose National Centre for Basic Sciences under Department of Science and Technology, Government of India. The authors thank the staff of IAO, Hanle and CREST, Hosakote, who made these observations possible. The facilities at IAO and CREST are operated by the Indian Institute of Astrophysics, Bangalore. We acknowledge the usage of the TIFR Near Infrared Spectrometer and Imager (TIRSPEC). SG is thankful to Joe Philip Ninan for helpful discussions and valuable suggestions about the data reduction on TIRSPEC-pipeline.








\appendix

\section{Some extra material}

\begin{table*}
	\centering
	\caption{ Measured Equivalent Widths of all the sample}
	\label{tab:all_measured_ews}
	\begin{tabular}{lccccccr} 
     \hline
	Star Names & Si I & CO1 & CO2 & Na I & Ca I & CO3 & CO4 \\
	\hline
 & & & \bf{TIRSPEC} : & & & & \\
HD54810 & 2.09 $\pm$0.24& 0.91 $\pm$0.66& 1.58 $\pm$0.23& 1.67 $\pm$0.36& 0.76 $\pm$0.44& 7.44 $\pm$2.24& 5.87 $\pm$1.13 \\
HD99283 & 2.54 $\pm$0.56& 0.82 $\pm$0.61& 1.35 $\pm$0.29& 0.90 $\pm$0.33& 1.60 $\pm$0.64& 8.19 $\pm$1.77& 4.54 $\pm$1.88\\
HD102224 & 2.31 $\pm$0.86& 0.84 $\pm$0.38& 1.78 $\pm$0.46& 0.76 $\pm$0.38& 2.02 $\pm$0.34& 11.82 $\pm$1.20& 8.68 $\pm$0.92\\
HD69994 & 2.52 $\pm$1.14& 1.15 $\pm$0.92& 1.66 $\pm$0.66& 1.57 $\pm$0.55& 1.44 $\pm$0.53& 10.32 $\pm$2.31& 6.19 $\pm$1.53\\
HD40657 & 2.70 $\pm$0.28& 1.64 $\pm$0.53& 2.71 $\pm$0.34& 1.40 $\pm$0.32& 0.24 $\pm$0.76& 11.63 $\pm$2.16& 7.48 $\pm$2.27 \\
HD85503 & 3.08 $\pm$0.47& 1.97 $\pm$0.65& 2.25 $\pm$0.60& 2.11 $\pm$1.38& 1.85 $\pm$1.16& 11.49 $\pm$2.74& 8.05 $\pm$1.91 \\
HD26846 & 3.13 $\pm$0.60& 1.81 $\pm$0.86& 2.73 $\pm$0.48& 2.73 $\pm$0.98& 1.23 $\pm$0.71& 11.70 $\pm$2.62& 7.00 $\pm$2.15 \\
HD30834 & 3.05 $\pm$0.38& 1.87 $\pm$0.77& 2.28 $\pm$0.35& 1.39 $\pm$0.41& 1.68 $\pm$0.51& 13.08 $\pm$0.95& 9.37 $\pm$1.71 \\
HD92523 & 2.50 $\pm$0.46& 1.79 $\pm$0.37& 2.84 $\pm$0.31& 1.30 $\pm$0.20& 1.85 $\pm$0.34& 14.74 $\pm$1.23& 9.52 $\pm$1.27 \\
HD97605 & 2.88 $\pm$0.92& 1.89 $\pm$1.86& 2.03 $\pm$0.73& 1.71 $\pm$0.38& 2.02 $\pm$0.70& 10.43 $\pm$1.36& 6.98 $\pm$1.71 \\
HD49161 & 3.11 $\pm$0.24& 1.97 $\pm$1.19& 2.70 $\pm$0.56& 2.64 $\pm$0.54& 2.45 $\pm$0.90& 12.43 $\pm$2.34& 11.82 $\pm$1.59 \\ 
HD70272 & 3.06 $\pm$0.49& 2.69 $\pm$0.28& 3.79 $\pm$0.62& 1.65 $\pm$0.32& 1.90 $\pm$0.40& 15.97 $\pm$1.75& 10.82 $\pm$1.32 \\
HD99167 & 2.85 $\pm$0.27& 3.56 $\pm$0.46& 4.23 $\pm$0.65& 2.80 $\pm$0.42& 2.19 $\pm$0.78& 17.11 $\pm$2.65& 12.73 $\pm$2.78 \\
HD83787 & 3.21 $\pm$0.67& 3.45 $\pm$0.60& 3.98 $\pm$1.15& 2.14 $\pm$0.71& 2.41 $\pm$0.51& 17.99 $\pm$1.97& 13.78 $\pm$2.12 \\
HD6953 & 3.22 $\pm$0.33& 3.61 $\pm$0.53& 3.87 $\pm$0.41& 2.02 $\pm$0.28& 3.05 $\pm$0.53& 16.10 $\pm$1.50& 9.79 $\pm$1.65 \\
HD6966 & 2.95 $\pm$0.77& 3.11 $\pm$0.44& 4.30 $\pm$0.96& 2.87 $\pm$0.36& 2.81 $\pm$0.45& 17.93 $\pm$1.73& 12.23 $\pm$2.18 \\
HD18760 & 3.04 $\pm$0.84& 2.25 $\pm$0.64& 4.82 $\pm$0.61& 2.16 $\pm$0.52& 1.80 $\pm$0.84& 14.36 $\pm$1.67& 11.17 $\pm$3.13 \\
HD38944 & 2.85 $\pm$0.76& 3.61 $\pm$1.10& 4.06 $\pm$0.31& 2.43 $\pm$0.55& 1.17 $\pm$1.02& 16.93 $\pm$2.42& 11.93 $\pm$1.10 \\
HD60522 & 3.60 $\pm$0.80& 3.66 $\pm$0.60& 4.71 $\pm$1.30& 2.76 $\pm$0.20& 3.26 $\pm$0.26& 17.51 $\pm$1.58& 10.72 $\pm$1.36 \\
HD216397 & 2.95 $\pm$0.58& 3.22 $\pm$1.11& 4.38 $\pm$0.87& 2.08 $\pm$0.41& 2.82 $\pm$0.71& 17.46 $\pm$1.88& 12.26 $\pm$1.81 \\
HD7158 & 2.83 $\pm$1.09& 4.41 $\pm$0.49& 5.18 $\pm$0.41& 1.94 $\pm$0.40& 2.34 $\pm$0.46& 18.18 $\pm$0.68& 12.06 $\pm$2.28 \\
HD82198 & 3.12 $\pm$0.83& 3.88 $\pm$0.46& 4.20 $\pm$0.40& 2.20 $\pm$0.46& 1.92 $\pm$0.33& 17.88 $\pm$2.15& 12.50 $\pm$1.83\\
HD218329 & 3.33 $\pm$0.90& 3.92 $\pm$0.65& 4.58 $\pm$0.46& 3.00 $\pm$0.99& 4.03 $\pm$0.71& 19.09 $\pm$4.36& 12.66 $\pm$3.53 \\
HD219215 & 2.91 $\pm$0.49& 4.02 $\pm$0.64& 5.16 $\pm$0.46& 2.68 $\pm$0.70& 4.06 $\pm$0.89& 19.71 $\pm$1.73& 13.02 $\pm$2.31 \\
HD119149 & 3.29 $\pm$0.62& 4.01 $\pm$0.60& 5.49 $\pm$0.77& 3.14 $\pm$0.47& 2.29 $\pm$1.20& 18.29 $\pm$3.03& 13.40 $\pm$3.19 \\
HD1013 & 3.13 $\pm$1.43& 4.18 $\pm$1.20& 5.30 $\pm$1.00& 2.52 $\pm$0.96& 3.34 $\pm$0.96& 19.86 $\pm$2.58& 12.79 $\pm$2.36 \\
HD33463 & 2.97 $\pm$1.57& 3.97 $\pm$1.46& 4.54 $\pm$1.09& 3.66 $\pm$0.77& 3.82 $\pm$1.60& 20.92 $\pm$2.07& 16.58 $\pm$2.52 \\
HD39732 & 2.77 $\pm$0.69& 4.31 $\pm$0.71& 5.32 $\pm$0.32& 2.04 $\pm$0.39& 1.88 $\pm$0.54& 21.57 $\pm$2.52& 13.75 $\pm$1.86 \\
HD43151 & 2.62 $\pm$0.18& 4.12 $\pm$0.41& 5.56 $\pm$0.78& 2.51 $\pm$0.26& 3.51 $\pm$0.38& 20.73 $\pm$1.24& 14.10 $\pm$1.80 \\
HD92620 & 2.91 $\pm$0.40& 4.09 $\pm$0.24& 5.42 $\pm$0.31& 2.66 $\pm$0.23& 2.99 $\pm$0.29& 20.58 $\pm$1.26& 14.50 $\pm$2.14 \\
HD115521 & 3.19 $\pm$0.51& 3.08 $\pm$0.66& 4.71 $\pm$0.57& 1.82 $\pm$0.25& 2.96 $\pm$0.47& 18.72 $\pm$1.77& 13.01 $\pm$1.30 \\
HD16058 & 2.83 $\pm$0.88& 4.72 $\pm$0.51& 6.57 $\pm$1.05& 2.72 $\pm$0.63& 3.23 $\pm$0.92& 22.17 $\pm$2.81& 12.35 $\pm$1.72 \\
HD28168 & 3.21 $\pm$1.00& 3.45 $\pm$1.61& 3.98 $\pm$0.74& 1.66 $\pm$1.01& 3.06 $\pm$1.36& 17.72 $\pm$10.61& 10.83 $\pm$2.02 \\
HD66875 & 2.75 $\pm$0.31& 4.11 $\pm$0.73& 5.60 $\pm$0.52& 2.69 $\pm$0.40& 3.72 $\pm$0.80& 22.45 $\pm$1.23& 14.07 $\pm$2.88 \\
HD99056 & 2.46 $\pm$0.50& 4.65 $\pm$0.67& 6.41 $\pm$0.45& 3.23 $\pm$0.44& 4.38 $\pm$1.09& 22.34 $\pm$2.38& 15.08 $\pm$3.13 \\
HD215953 & 3.12 $\pm$0.47& 4.68 $\pm$1.53& 5.72 $\pm$0.87& 3.38 $\pm$1.26& 4.34 $\pm$0.98& 21.82 $\pm$2.74& 14.04 $\pm$1.99 \\
HD223637 & 3.20 $\pm$0.52& 3.48 $\pm$0.73& 4.95 $\pm$0.34& 2.25 $\pm$0.76& 2.42 $\pm$0.82& 20.57 $\pm$2.78& 11.50 $\pm$2.15 \\
HD25921 & 3.26 $\pm$0.75& 3.80 $\pm$0.98& 6.16 $\pm$1.28& 3.25 $\pm$0.49& 3.76 $\pm$0.63& 18.34 $\pm$2.13& 12.78 $\pm$2.14 \\
HD33861 & 3.43 $\pm$0.90& 4.54 $\pm$1.14& 6.40 $\pm$0.50& 2.71 $\pm$0.41& 2.93 $\pm$0.36& 25.16 $\pm$1.58& 16.41 $\pm$2.10 \\
HD224062 & 2.66 $\pm$0.17& 4.09 $\pm$0.66& 5.53 $\pm$0.51& 2.20 $\pm$0.61& 2.96 $\pm$0.97& 21.41 $\pm$1.86& 12.01 $\pm$2.33 \\
HD5316 & 3.13 $\pm$0.40& 4.18 $\pm$0.44& 5.30 $\pm$1.27& 2.53 $\pm$0.61& 4.10 $\pm$1.00& 21.28 $\pm$1.51& 13.79 $\pm$2.55 \\
HD34269 & 2.62 $\pm$1.13& 4.33 $\pm$0.58& 5.88 $\pm$0.70& 2.91 $\pm$0.59& 4.81 $\pm$1.35& 22.61 $\pm$3.61& 14.24 $\pm$1.91 \\
HD64052 & 2.99 $\pm$1.49& 4.84 $\pm$1.17& 6.80 $\pm$0.69& 4.35 $\pm$0.66& 3.10 $\pm$1.49& 21.16 $\pm$3.74& 16.43 $\pm$3.21 \\
HD81028 & 2.70 $\pm$0.48& 3.98 $\pm$0.51& 5.41 $\pm$0.27& 2.02 $\pm$0.24& 2.88 $\pm$0.42& 20.13 $\pm$1.39& 12.84 $\pm$2.13 \\
HD206632 & 2.98 $\pm$0.91& 4.53 $\pm$0.59& 6.47 $\pm$0.60& 4.44 $\pm$0.73& 4.72 $\pm$1.64& 22.44 $\pm$3.05& 14.41 $\pm$3.79 \\
HD16896 & 2.67 $\pm$0.46& 3.68 $\pm$0.75& 5.83 $\pm$1.08& 3.14 $\pm$0.63& 4.61 $\pm$1.33& 20.86 $\pm$2.45& 15.38 $\pm$2.37 \\
HD17491 & 3.21 $\pm$1.36& 3.64 $\pm$0.72& 5.98 $\pm$0.72& 3.09 $\pm$0.55& 3.14 $\pm$0.44& 20.25 $\pm$2.52& 14.86 $\pm$2.12 \\
HD17895 & 2.95 $\pm$1.33& 3.14 $\pm$0.79& 6.18 $\pm$0.50& 3.50 $\pm$0.53& 4.07 $\pm$0.47& 19.74 $\pm$1.48& 14.21 $\pm$1.95 \\
HD22689 & 2.82 $\pm$0.41& 4.21 $\pm$0.77& 6.55 $\pm$0.86& 2.25 $\pm$0.85& 3.87 $\pm$0.57& 23.24 $\pm$3.49& 12.41 $\pm$2.22 \\
HD26234 & 2.83 $\pm$0.46& 4.72 $\pm$1.44& 6.57 $\pm$0.78& 2.72 $\pm$0.44& 3.23 $\pm$0.59& 22.17 $\pm$2.69& 12.35 $\pm$2.21 \\
HD39983 & 2.56 $\pm$0.48& 3.38 $\pm$0.36& 5.59 $\pm$0.57& 2.25 $\pm$0.40& 3.03 $\pm$0.48& 19.65 $\pm$1.76& 13.37 $\pm$1.77 \\
HD46421 & 2.60 $\pm$0.55& 4.27 $\pm$0.65& 6.54 $\pm$1.12& 3.10 $\pm$0.25& 2.94 $\pm$0.80& 23.37 $\pm$2.53& 15.00 $\pm$1.90 \\
HD66175 & 3.15 $\pm$0.29& 5.08 $\pm$0.80& 7.45 $\pm$0.56& 2.56 $\pm$0.31& 4.15 $\pm$0.32& 24.02 $\pm$1.21& 15.97 $\pm$1.42 \\
HD103681 & 3.45 $\pm$0.68& 5.30 $\pm$0.48& 7.37 $\pm$0.37& 2.30 $\pm$0.31& 3.38 $\pm$0.30& 24.75 $\pm$0.63& 16.23 $\pm$1.43 \\
HD105266 & 2.65 $\pm$0.32& 5.43 $\pm$0.55& 7.00 $\pm$0.93& 4.44 $\pm$0.60& 3.24 $\pm$1.16& 23.39 $\pm$2.98& 17.58 $\pm$3.42 \\
HD64657 & 2.76 $\pm$0.42& 4.32 $\pm$0.21& 5.89 $\pm$1.33& 2.61 $\pm$1.03& 4.22 $\pm$1.24& 22.67 $\pm$2.52& 15.67 $\pm$3.42 \\
HD65183 & 2.99 $\pm$0.88& 4.14 $\pm$0.94& 5.77 $\pm$0.63& 2.29 $\pm$0.42& 3.25 $\pm$1.46& 21.84 $\pm$2.95& 14.18 $\pm$1.67 \\
HD223608 & 2.98 $\pm$0.69& 4.16 $\pm$0.42& 6.24 $\pm$0.34& 2.66 $\pm$0.58& 3.62 $\pm$0.72& 23.41 $\pm$1.12& 15.44 $\pm$1.63 \\
HD7861 & 2.80 $\pm$1.48& 4.22 $\pm$0.75& 6.26 $\pm$1.48& 2.69 $\pm$0.49& 4.58 $\pm$0.95& 19.80 $\pm$1.99& 14.82 $\pm$1.51 \\
HD18191 & 3.32 $\pm$0.93& 4.01 $\pm$0.49& 6.65 $\pm$1.04& 2.74 $\pm$0.64& 4.52 $\pm$0.73& 21.54 $\pm$1.81& 15.61 $\pm$2.09 \\
HD27957 & 2.69 $\pm$1.04& 3.66 $\pm$0.54& 6.28 $\pm$0.63& 3.25 $\pm$0.36& 4.38 $\pm$0.29& 21.59 $\pm$1.51& 14.78 $\pm$2.25 \\
HD70421 & 2.54 $\pm$0.52& 4.39 $\pm$0.98& 6.10 $\pm$1.04& 2.62 $\pm$0.45& 4.29 $\pm$0.50& 22.17 $\pm$1.78& 14.34 $\pm$2.17 \\
HD73844 & 2.92 $\pm$1.08& 4.59 $\pm$0.85& 7.28 $\pm$1.15& 2.17 $\pm$0.77& 4.72 $\pm$1.10& 24.14 $\pm$2.50& 15.35 $\pm$1.60 \\

		\hline
	\end{tabular}
\end{table*}

\begin{table*}
	\centering
	\contcaption{}
	\begin{tabular}{lccccccr} 
     \hline
		Star Names & Na I & CO1 & CO2 & Na I & Ca I & CO3 & CO4 \\
		\hline

HIP44601 & 2.66 $\pm$0.45& 4.41 $\pm$0.52& 6.15 $\pm$0.71& 2.86 $\pm$0.25& 4.49 $\pm$0.87& 25.44 $\pm$2.06& 16.42 $\pm$1.64 \\
HIC55173 & 2.65 $\pm$0.62& 4.70 $\pm$0.51& 6.36 $\pm$0.85& 3.23 $\pm$0.65& 4.19 $\pm$0.58& 24.83 $\pm$2.23& 16.08 $\pm$2.93 \\             
HIP57504 & 2.62 $\pm$1.09& 4.68 $\pm$0.61& 6.19 $\pm$0.66& 2.63 $\pm$0.29& 2.91 $\pm$0.27& 20.64 $\pm$1.00& 13.80 $\pm$2.25 \\
HD115322 & 3.17 $\pm$0.59& 3.98 $\pm$0.66& 5.43 $\pm$0.66& 2.43 $\pm$0.83& 3.41 $\pm$0.81& 18.47 $\pm$3.15& 14.24 $\pm$1.42 \\
HD203378 & 2.98 $\pm$1.41& 4.53 $\pm$0.85& 6.47 $\pm$1.60& 3.62 $\pm$0.71& 4.69 $\pm$1.06& 22.09 $\pm$2.63& 14.46 $\pm$2.81 \\
HD43635 & 2.84 $\pm$0.57& 4.15 $\pm$0.51& 6.20 $\pm$1.47& 2.56 $\pm$0.69& 4.76 $\pm$0.56& 24.43 $\pm$2.26& 14.47 $\pm$2.51 \\
HIC51353 & 2.80 $\pm$0.70& 4.70 $\pm$2.16& 6.96 $\pm$0.63& 3.64 $\pm$1.10& 3.68 $\pm$0.72& 22.42 $\pm$1.35& 17.20 $\pm$2.52 \\
HIC68357 & 2.67 $\pm$0.96& 4.82 $\pm$1.14& 6.16 $\pm$1.45& 3.67 $\pm$0.79& 3.01 $\pm$1.23& 22.39 $\pm$3.62& 16.25 $\pm$2.94 \\
HD141265 & 2.67 $\pm$0.64& 5.49 $\pm$0.80& 6.95 $\pm$1.39& 2.39 $\pm$0.66& 3.96 $\pm$0.78& 23.48 $\pm$1.62& 12.89 $\pm$2.72 \\
 & & & \bf{SpeX} : & & & & \\
HD100006 & 2.79 $\pm$0.60& 1.12 $\pm$0.67& 1.29 $\pm$0.33& 1.06 $\pm$0.21& 0.97 $\pm$0.15& 8.45 $\pm$1.10& 5.84 $\pm$1.48 \\
HD9852 & 2.92 $\pm$0.52& 1.92 $\pm$0.82& 1.95 $\pm$0.29& 1.53 $\pm$0.24& 1.56 $\pm$0.25& 11.80 $\pm$1.47& 8.25 $\pm$1.36 \\
HD25975 & 2.57 $\pm$0.33& 2.01 $\pm$0.68& 1.11 $\pm$0.27& 1.30 $\pm$0.24& 1.24 $\pm$0.22& 6.09 $\pm$0.89& 4.40 $\pm$1.05 \\
HD36134 & 2.71 $\pm$0.21& 1.92 $\pm$0.62& 1.54 $\pm$0.27& 1.22 $\pm$0.20& 1.36 $\pm$0.21& 9.54 $\pm$1.64& 6.79 $\pm$1.28 \\
HD91810 & 3.06 $\pm$0.30& 2.29 $\pm$0.55& 2.04 $\pm$0.25& 1.64 $\pm$0.28& 1.71 $\pm$0.50& 10.76 $\pm$1.48& 7.57 $\pm$1.67 \\
HD124897 & 2.97 $\pm$0.44& 2.35 $\pm$0.43& 2.36 $\pm$0.37& 1.22 $\pm$0.16& 1.60 $\pm$0.45& 13.39 $\pm$1.93& 9.48 $\pm$1.73 \\
HD137759 & 2.85 $\pm$0.59& 2.18 $\pm$0.42& 1.91 $\pm$0.29& 1.62 $\pm$0.25& 1.77 $\pm$0.17& 10.30 $\pm$1.47& 6.86 $\pm$1.19 \\
HD132935 & 2.81 $\pm$0.49& 2.12 $\pm$0.35& 2.33 $\pm$0.41& 1.31 $\pm$0.19& 1.60 $\pm$0.36& 13.34 $\pm$1.53& 10.10 $\pm$2.15 \\
HD2901 & 2.81 $\pm$0.45& 2.29 $\pm$0.47& 2.31 $\pm$0.24& 1.47 $\pm$0.28& 1.49 $\pm$0.29& 12.69 $\pm$1.41& 9.13 $\pm$1.91 \\
HD221246 & 3.17 $\pm$0.90& 2.80 $\pm$0.67& 3.19 $\pm$0.46& 2.08 $\pm$0.26& 2.16 $\pm$0.28& 14.71 $\pm$1.70& 10.87 $\pm$2.05 \\
HD178208 & 3.18 $\pm$0.49& 2.54 $\pm$0.92& 2.92 $\pm$0.36& 2.23 $\pm$0.33& 1.99 $\pm$0.40& 13.67 $\pm$1.44& 9.71 $\pm$1.85 \\
HD35620 & 2.91 $\pm$0.36& 2.72 $\pm$0.90& 3.29 $\pm$0.41& 2.17 $\pm$0.49& 2.07 $\pm$0.39& 15.39 $\pm$1.52& 10.90 $\pm$2.01 \\
HD99998 & 3.19 $\pm$0.75& 3.42 $\pm$0.46& 3.67 $\pm$0.58& 2.00 $\pm$0.43& 1.89 $\pm$0.34& 15.64 $\pm$1.45& 11.70 $\pm$2.30 \\
HD114960 & 3.10 $\pm$0.68& 3.04 $\pm$0.61& 3.69 $\pm$0.54& 2.55 $\pm$0.52& 2.41 $\pm$0.40& 14.91 $\pm$2.18& 10.44 $\pm$2.05 \\
HD207991 & 3.15 $\pm$0.71& 3.02 $\pm$0.56& 3.56 $\pm$0.53& 1.99 $\pm$0.27& 1.96 $\pm$0.34& 15.99 $\pm$1.61& 11.88 $\pm$1.52 \\
HD181596 & 3.24 $\pm$0.74& 3.17 $\pm$0.48& 4.27 $\pm$0.48& 2.57 $\pm$0.37& 2.30 $\pm$0.30& 17.20 $\pm$1.78& 12.67 $\pm$2.58 \\
HD120477 & 3.30 $\pm$0.73& 3.98 $\pm$0.47& 4.55 $\pm$0.55& 2.55 $\pm$0.53& 2.54 $\pm$0.40& 17.48 $\pm$1.54& 11.89 $\pm$2.29 \\
HD3346 & 3.21 $\pm$0.40& 3.83 $\pm$0.36& 4.40 $\pm$0.57& 2.43 $\pm$0.47& 2.19 $\pm$0.25& 17.89 $\pm$2.15& 12.54 $\pm$2.21 \\
HD194193 & 3.20 $\pm$0.35& 3.51 $\pm$0.45& 4.36 $\pm$0.81& 2.43 $\pm$0.36& 2.47 $\pm$0.42& 17.80 $\pm$1.90& 12.61 $\pm$2.86 \\
HD213893 & 3.35 $\pm$0.88& 3.43 $\pm$0.45& 4.36 $\pm$0.69& 2.23 $\pm$0.42& 2.12 $\pm$0.32& 16.96 $\pm$2.22& 12.66 $\pm$2.25 \\
HD204724 & 2.86 $\pm$0.70& 3.07 $\pm$0.26& 4.21 $\pm$0.73& 3.17 $\pm$0.48& 3.14 $\pm$0.53& 18.68 $\pm$2.81& 13.08 $\pm$3.02 \\
HD120052 & 3.32 $\pm$1.63& 3.28 $\pm$0.42& 4.45 $\pm$0.70& 2.37 $\pm$0.25& 2.33 $\pm$0.32& 18.33 $\pm$2.35& 13.26 $\pm$2.20 \\
HD219734 & 3.21 $\pm$1.69& 4.13 $\pm$0.45& 4.82 $\pm$0.51& 2.96 $\pm$0.42& 2.75 $\pm$0.46& 18.87 $\pm$2.32& 14.28 $\pm$2.25 \\
HD39045 & 2.80 $\pm$1.15& 5.07 $\pm$0.60& 5.24 $\pm$0.61& 2.88 $\pm$0.81& 2.54 $\pm$0.41& 19.35 $\pm$2.23& 13.86 $\pm$2.84 \\
HD28487 & 2.65 $\pm$0.60& 3.53 $\pm$0.56& 5.27 $\pm$0.90& 2.80 $\pm$0.40& 2.74 $\pm$0.55& 20.04 $\pm$2.45& 14.09 $\pm$3.42 \\
HD4408 & 2.98 $\pm$0.36& 4.20 $\pm$0.84& 5.50 $\pm$0.81& 3.49 $\pm$0.55& 3.20 $\pm$0.36& 20.30 $\pm$2.35& 13.84 $\pm$2.81 \\
HD204585 & 3.04 $\pm$0.36& 4.56 $\pm$0.91& 5.92 $\pm$0.84& 4.10 $\pm$0.60& 3.70 $\pm$0.48& 21.89 $\pm$2.54& 15.61 $\pm$3.39 \\
HD27598 & 2.54 $\pm$0.43& 3.99 $\pm$0.84& 5.09 $\pm$0.59& 2.70 $\pm$0.33& 2.66 $\pm$0.39& 20.60 $\pm$2.43& 14.64 $\pm$2.31 \\
HD19058 & 3.06 $\pm$0.76& 4.13 $\pm$0.52& 5.58 $\pm$0.70& 3.39 $\pm$0.44& 3.01 $\pm$0.33& 19.13 $\pm$2.40& 13.43 $\pm$2.33 \\
HD214665 & 2.94 $\pm$0.98& 4.54 $\pm$0.53& 5.77 $\pm$0.72& 3.37 $\pm$0.63& 3.35 $\pm$0.54& 21.17 $\pm$2.73& 14.35 $\pm$2.19 \\
HD175865 & 2.74 $\pm$1.23& 5.39 $\pm$0.70& 6.32 $\pm$0.79& 3.97 $\pm$0.62& 3.53 $\pm$0.54& 21.75 $\pm$2.24& 15.50 $\pm$2.69 \\
HD94705 & 2.19 $\pm$1.03& 4.75 $\pm$0.74& 5.44 $\pm$1.18& 4.16 $\pm$0.91& 3.61 $\pm$0.64& 20.33 $\pm$2.55& 13.47 $\pm$2.59 \\
HD196610 & 2.67 $\pm$0.74& 5.03 $\pm$0.71& 6.43 $\pm$0.79& 3.49 $\pm$0.50& 2.99 $\pm$0.69& 23.51 $\pm$2.80& 17.24 $\pm$2.77 \\
HD108849 & 1.77 $\pm$0.67& 4.45 $\pm$0.71& 6.06 $\pm$2.82& 3.98 $\pm$0.42& 2.88 $\pm$0.51& 22.90 $\pm$2.39& 15.39 $\pm$2.67 \\
BRI2339-0447 & 1.01 $\pm$0.45& 1.58 $\pm$1.21& 3.32 $\pm$0.87& 1.90 $\pm$0.16& -0.35 $\pm$0.40& 18.36 $\pm$1.44& 10.52 $\pm$1.73 \\

		\hline
	\end{tabular}	

      \small
      The Table~\ref{tab:all_measured_ews} is available in its entirety in the electronic version of the journal.
\end{table*}


\bsp	
\label{lastpage}
\end{document}